\documentclass[12pt]{article}

\usepackage{amsmath, amssymb, hyperref}
\usepackage{epsfig,amsfonts,verbatim,mathrsfs}
\usepackage{dcolumn,bm,multirow}

\newcommand{\beq}{\begin{eqnarray}}
\newcommand{\eeq}{\end{eqnarray}}
\newcommand{\bea}{\begin{eqnarray}}
\newcommand{\eea}{\end{eqnarray}}
\newcommand{\lrf}[2]{\left(\frac{#1}{#2}\right)}
\newcommand{\met}{\,/\hspace{-0.25cm}E_T}

\newcommand{\nnmb}{\nonumber}

\newcommand{\tev}{\, {\rm TeV}}
\newcommand{\gev}{\, {\rm GeV}}

\newcommand{\ccdot}{\!\cdot\!}
\newcommand{\til}{\tilde}

\newcommand{\MET}{\ensuremath{{\not\mathrel{E}}_T}}
\newcommand{\mstop}{m_{\til t}}
\newcommand{\mneut}{m_{\chi_{1}^{0}}}
\newcommand{\ifb}{fb^{-1}}
\newcommand{\newc}{\newcommand}
\newc{\vu}{v_u}
\newc{\vd}{v_d}
\newc{\yu}{{Y}_u}
\newc{\yd}{{Y}_d}
\newc{\ye}{{Y}_e}
\newc{\Au}{{a}_u}
\newc{\Ad}{{a}_d}
\newc{\Ae}{{a}_e}

\begin{document}

\begin{titlepage}
\noindent
\vspace{1cm}
\begin{center}
  \begin{Large}
    \begin{bf}
Very Light Scalar Top Quarks at the LHC
     \end{bf}
  \end{Large}
\end{center}
\vspace{0.2cm}

\begin{center}

\begin{large}
Karol Krizka$^{(a,b,c)}$, Abhishek Kumar$^{(b)}$,\\
David E. Morrissey$^{(b)}$
\end{large}
\vspace{1cm}\\
  \begin{it}
(a) Department of Physics, Simon Fraser University, Burnaby, BC V5A~1S6, Canada\\
(b) TRIUMF, 4004 Wesbrook Mall, Vancouver, BC V6T~2A3, Canada\\
(c) Enrico Fermi Institute and Department of Physics,
University of Chicago, 5620 S. Ellis Avenue, Chicago, IL 60637, USA
%
%
\vspace{0.2cm}\\
email: kkrizka@uchicago.edu, abhishek@triumf.ca, dmorri@triumf.ca
\vspace{0.2cm}
\end{it}

\end{center}

\center{\today}


\begin{abstract}

  A very light scalar top (stop) superpartner is motivated by naturalness 
and electroweak baryogenesis.  When the mass of the stop 
is less than the sum of the masses of the top quark and the lightest neutralino 
superpartner $\chi_1^0$, as well as the of the masses of the lightest 
chargino and the bottom quark, the dominant decay channels of the stop 
will be three-body~(3B: $bW^+\chi_1^0$), four-body~(4B: $bf\bar{f}'\chi_1^0$), 
or flavour violating~(FV: $c\chi_1^0$).  
In this work, we investigate the direct and indirect constraints on a light
stop, we compute the relative decay branching fractions to these channels,
and we study the sensitivity of existing LHC searches to each of them.

\end{abstract}

\end{titlepage}

\setcounter{page}{2}


\section{Introduction\label{sec:intro}}

  Supersymmetry can protect the scale of electroweak symmetry
breaking from destabilising quantum corrections provided the masses of the
Standard Model~(SM) superpartners are not too large~\cite{Martin:1997ns}.  
The most important states in this regard are those that couple 
the most strongly to the Higgs fields, namely the scalar top quarks (stops).  
To give the desired protection, the two stop mass eigenstates should 
be considerably lighter than a TeV~\cite{Dimopoulos:1995mi}.

  A light stop can also play an important role in cosmology.
The baryon asymmetry can be created within the minimal supersymmetric
standard model~(MSSM) by the process of electroweak baryogenesis~(EWBG)
if at least one of the stops is lighter than the 
top quark~\cite{Carena:2008vj,Laine:2012jy,Morrissey:2012db}.
A light stop can help to produce the observed dark matter
relic density as well.  When the lightest superpartner~(LSP) is a
mostly-Bino neutralino, the mechanism of thermal freeze-out tends
to create too much dark matter.  However, a light stop that is close
in mass to the neutralino LSP can reduce its thermal abundance
to the observed value by coannihilation~\cite{Boehm:1999bj,Ellis:2001nx,Balazs:2004bu,Balazs:2004ae}.  

  For these many reasons, an intense search for light stops is underway at
the Large Hadron Collider~(LHC), and a broad range of theoretical studies have
been performed recently to find the most promising search channels~\cite{Kats:2011it,Papucci:2011wy,Essig:2011qg,Plehn:2012pr,Alves:2012ft,Han:2012fw,Kaplan:2012gd,Cao:2012rz,Dutta:2012kx,Kilic:2012kw}.  
The two decay channels that have been studied in the most detail are
\beq
\tilde{t}_1\to t\chi_1^0,~~~~~\tilde{t}_1\to b\chi_1^+~~~~~(\text{2B}) \ ,
\eeq
where $\chi_1^0$ is the lightest neutralino and $\chi_1^+$ is the lightest
chargino.  When the stop is too light to decay in these channels,
the dominant decay modes can involve three- or four-body final states 
such as~\cite{Bigi:1985aq,Hikasa:1987db,Boehm:1999tr}
\beq
\tilde{t}\to b\chi_1^0W^{+(*)}~~~~~~(\text{3B, 4B}) \ ,
\eeq
or have flavour violation~\cite{Bigi:1985aq,Hikasa:1987db}, 
\beq
\tilde{t}\to c\chi_1^0~~~~~(\text{FV}) \ .
\eeq
Of these, the two-body FV mode has been studied 
the most extensively~\cite{Carena:2008mj,Demina:1999ty,Bi:2011ha,Ajaib:2011hs,He:2011tp,Yu:2012kj,Drees:2012dd,Kraml:2005kb,Bornhauser:2010mw,Martin:2008aw,Das:2001kd,Choudhury:2012kn,Ghosh:2012ud,Belanger:2012mk,Dreiner:2012sh}, and is typically assumed to be the
exclusive decay channel when the three-body decay is kinematically forbidden. 

  In this work we study the LHC collider signals of a very light stop, 
with $m_{\tilde{t}_1} \leq m_t + m_{\chi_1^0}$, in a comprehensive way.  
We place an emphasis on the scenario where only the 4B and FV decay channels 
are open.  The branching ratios of these modes depend on the 
underlying flavour structure of the theory, and we make
the assumption that it is governed by Minimal Flavour Violation~(MFV).  
In this case, we find that the $4B$ channel is frequently the dominant 
one despite its suppression by the large final-state phase space.  
To our knowledge, the LHC signals of a light stop decaying 
in this way have not been studied in detail.

  The outline of this paper is as follows.  We begin in Sec.~\ref{sec:stop}
by presenting the relevant stop interactions and the corresponding
mass matrix, as well as our assumptions about MFV flavour mixing.
Next, in Sec.~\ref{sec:limits} we investigate the indirect limits 
on very light stops from precision electroweak and flavour measurements. 
With these bounds in hand, we study the dominant 
decay channels of the light stop in Sec.~\ref{sec:decay}.  
We then apply these results to make estimates of the sensitivity 
of existing LHC searches to a light stop.
Finally, Sec.~\ref{sec:conc} is reserved for our conclusions.
Some technical details can be found in the Appendix~\ref{sec:mfv}.

\section{Stop Parameters\label{sec:stop}}

  To begin, we define our parameters and specify our assumptions
about the flavour structure of the theory.  In the absence of
flavour mixing, the tree-level stop mass matrix is~\cite{Martin:1997ns}
\beq
\mathcal{M}_{\tilde{t}}^2 = \left( \begin{array}{cc}
m_{Q_3}^{2} + (\frac{1}{2}-\frac{2}{3} s_{W}^{2})c_{2\beta}m_{Z}^{2}  + m_{t}^{2} & m_{t} X_{t}^* \\
m_{t} X_{t} & m_{U^c_3}^{2} + \frac{2}{3} s_{W}^{2}c_{2\beta}m_{Z}^{2}  
+ m_{t}^{2} 
\end{array} \right) \ ,
\eeq 
with $m_{Q_3}^2$ and $m_{U^c_3}^2$ the scalar soft squared masses, 
$m_t$ mass of the top quark, $\tan\beta$ the ratio of Higgs 
vacuum expectation values~(VEVs), and 
\beq
X_{t} = A_{t} - \mu^*/\tan\beta
\eeq
parametrizes the left-right~(LR) stop mixing.  A light stop mass eigenstate 
arises for small soft mass parameters or large LR mixing. 

  Flavour mixing within the Yukawa couplings of the SM translates 
into some degree of flavour mixing in its supersymmetric extension.
Additional sources of flavour mixing can arise from soft supersymmetry
breaking parameters, although these are strongly constrained by 
experiment~\cite{Altmannshofer:2009ne,Hurth:2012jn}.  
In this work we will assume a specific flavour structure for 
the soft terms that is relatively consistent with existing flavour bounds, 
namely Minimal Flavour Violation~(MFV)~\cite{D'Ambrosio:2002ex}. 

The structure of MFV is based on the observation that the SM has 
a global $SU(3)^{5}$ flavour symmetry in the absence of Yukawa couplings. 
By promoting the Yukawas to spurions that transform under the global 
flavour group, the SM is made invariant under the symmetry. 
The flavour group is then broken only by the background values of the spurions. 
As applied to theories of new physics beyond the SM, MFV requires that 
this flavour structure continue to hold true,
with any new sources of flavour mixing proportional to the Yukawa spurions.

  For the MSSM, the MFV hypothesis restricts the flavour structure
of the soft parameters.  The soft scalar mass-squareds
become $6\times 6$ matrices.  To leading order in the Yukawa expansion 
they are given by~\cite{D'Ambrosio:2002ex,Colangelo:2008qp}
\bea
{m}_{Q}^{2} &=& \til m^{2} (a_{1}\mathbb{I} + b_{1} \yu \yu^{\dag} 
+ b_{2} \yd \yd^{\dag}
+ b_3\yd\yd^{\dag}\yu\yu^{\dag}
+b_4\yu\yu^{\dag}\yd\yd^{\dag}) \label{mfv1}\\
m_{U^c}^{2} &=& \til m^{2} (a_{2}\mathbb{I} + b_{5} \yu^{\dag} \yu 
+ c_{1} \yu^{\dag}\yd \yd^{\dag}\yu) \label{mfv2}\\ 
m_{D^c}^2 &=& \til m^{2} (a_{3}\mathbb{I} + b_{6} \yd^{\dag}\yd ) \ .
\label{mfv3}
\eea
Similarly, the squark trilinear couplings are
\bea
     a_{u} &=& A (a_{4} {\bf 1} + b_{7} \yd\yd^{\dag})\yu
\label{mfv4}\\
     a_{d} &=& A (a_{5} {\bf 1} + b_{8} \yu\yu^{\dag})\yd \ .
\label{mfv5}
\eea
The coefficients $a_{i}$, $b_{i}$, and $c_{i}$ parametrize the
expansion and are expected to be on the order of unity or less.
Their values may be restricted within specific theories of flavour,
but we treat them as free parameters at the superpartner mass 
scale.\footnote{This is consistent with but more general than the commonly-used
assumption that the flavour mixing vanishes at some high-energy scale.}  
See Appendix~\ref{sec:mfv} for a more complete description of our conventions 
for flavour and the soft masses.

  To compute flavour-violating stop decays, it is convenient to
choose a specific basis for the Yukawa matrices where the up-Yukawas
are diagonal,
\beq
\yu = \lambda_{u} \qquad \qquad \qquad \yd = V \lambda_{d} \ ,
\eeq
where $V$ is the CKM matrix, and $\lambda_{u,d}$ are the diagonal 
up- and down-Yukawa matrices. 

  For our study we choose a set of fiducial soft parameters that allows
us to isolate a light stop state.  We take $\tilde{m} = 1500\,\gev$
and $a_1=a_2=a_3=1$, which has the effect of decoupling the first- and
second-generation squarks, and we choose similar soft masses
for the sleptons.  Much smaller values of $(m_Q^2)_{33}$
and $(m_{U^c}^2)_{33}$ are obtained by suitable choices of $b_1$ and $b_5$,
and these translate into one or more light squark states.  
Since these states only contain very small admixtures of the first- 
and second-generation squarks, we identify them with a pair of stops 
and a (mainly left-handed) sbottom.  The rest of the $b_i$ and $c_i$ 
parameters are scanned over the range $[-1,1]$.  
We also set the gluino soft mass to $M_3 = 1000\,\gev$, 
but we consider a range of values for the
electroweak gaugino soft masses $M_1$ and $M_2$ and the Higgsino
mass $\mu$.  In implementing these values within the spectrum
generators SuSpect 2.41~\cite{Djouadi:2002ze} 
and SoftSUSY 3.3.5~\cite{Allanach:2001kg}, 
we input these parameters as running values defined at 
$M_S = \sqrt{m_tm_Q}$.
Finally, we consider the specific representative cases of 
$\tan\beta = 10,\,30$.

\section{Limits on a Light Stop\label{sec:limits}}

  Stops have been searched for extensively at the Tevatron and
the LHC.  A light stop can also influence a number of SM observables by its
contributions in loops.  We review here the existing limits on stops
from direct collider searches and we study the indirect limits on a light
stop from precision electroweak measurements and flavour violation.
We also comment on the constraints imposed by the rates of Higgs 
production and decay.

\subsection{Direct Collider Searches} 

  The ATLAS and CMS collaborations at the CERN LHC have succeeded in
confirming the SM to such an extent that squarks of the first and 
second generations are constrained to be heavier than a TeV~\cite{A2,:2012mfa}.
More recently, these collaborations have turned their attention
to investigating the squarks of the third generation created both
directly and indirectly.  Recent studies rule out direct 
stop pair production for stop masses
up to nearly $600\,\gev$ under the assumption that the only decay
is $\tilde{t}\to t\chi_1^0$
with $m_{\chi_1^0} \to 0$ and $\chi_1^0$ a stable LSP~\cite{Ast1,Ast2}.  
The limits weaken for larger values of the LSP mass, but remain in the range
of many hundreds of GeV.  Searches for direct sbottom pair production
with $\tilde{b}\to b\chi_1^0$ yield similar bounds on the lightest 
sbottom mass~\cite{Asbot}.
Related searches for direct stop production with the decay 
$\tilde{t}\to b\chi_1^+$ together with $\chi_1^+\to W^{+(*)}\chi_1^0$ 
also give exclusions as large as $m_{\tilde{t}} > 450\,\gev$~\cite{Ast1,Ast2}.  
Finally, light stops or sbottoms can 
be created in gluino decays.  Assuming one or the other of the 
topologies $\tilde{g}\to t\bar{t}\chi_1^0$ with $\tilde{t}\to t\chi_1^0$
or $\tilde{g}\to b\tilde{b}$ with $\tilde{b}\to b\chi_1^0$,
the constraint on the gluino mass extends up to  
$m_{\tilde{g}} = 1.3\,\tev$~\cite{Astglu,Cstglu}.  

    Collider limits on stops and sbottoms can be much weaker when
they decay in other ways, such as the 3B, 4B, and FV channels.    
Searches at LEP rule out a stop decaying in the FV channel up to about 
$100\,\gev$~\cite{Abbiendi:2002mp}, and we expect similar limits to hold
for the 3B and 4B decays.  At the Tevatron,
specific searches for a light stop in the FV mode give limits 
up to $m_{\tilde{t}}\simeq 180\,\gev$, but these limits disappear 
when the stop becomes moderately degenerate with 
the LSP~\cite{Aaltonen:2012tq,Abazov:2008rc}.  
Almost no limit is expected when the 3B or 4B channels 
are dominant~\cite{Demina:1999ty}.   
The LHC collaborations have not performed dedicated searches for a stop
in the 3B, 4B, and FV modes.  The goal of this work is deduce
the sensitivity of existing LHC searches to these stop channels.

\subsection{Indirect Bounds from Electroweak and Flavour}

  The most important effect of a light stop on precision electroweak 
observables is to modify the prediction for the mass of the $W$ boson,
corresponding to a shift in $\Delta\rho$ (or $T$ in the $STU$ parametrization
of oblique parameters~\cite{Peskin:1991sw}).  
This effect has been studied recently
in Refs.~\cite{Lee:2012sy,Barger:2012hr,Espinosa:2012in} 
in light of the improved determination
of $M_W$~\cite{Group:2012gb} together with evidence for a Higgs boson
of mass $m_h=125\,\gev$~\cite{:2012gk,:2012gu}.  These results imply 
$\Delta\rho = (4.2\pm 2.7)\times 10^{-4}$~\cite{Barger:2012hr,
Espinosa:2012in}.  

  The change in $\Delta\rho$ due to light stops is given
by~\cite{Heinemeyer:2004gx}
\beq
\Delta\rho \simeq \frac{3G_F}{8\sqrt{2}\pi^2}\,\left[
-s_{\tilde{t}}^2c_{\tilde{t}}^2F_0(m_{\tilde{t}_1}^2,m_{\tilde{t}_2}^2)
+c_{\tilde{t}}^2F_0(m_{\tilde{t}_1}^2,m_{\tilde{b}_L}^2)
+s_{\tilde{t}}^2F_0(m_{\tilde{t}_2}^2,m_{\tilde{b}_L}^2)\right] \ ,
\eeq
where $\theta_{\tilde{t}}$ is the stop mixing angle and
$F_0(x,y)$ is a loop function,
\beq
F_0(x,y) = x+y - \frac{2xy}{x-y}\ln\lrf{x}{y} \ .
\eeq
We apply this formula in the present case where we have
squark flavour mixing dictated by the MFV paradigm.  Since
the inter-generational mixing is numerically small, we are able
to identify specific squark mass eigenstates with ${\tilde{t}_{1}}$,
${\tilde{t}_{2}}$, and ${\tilde{b}_{L}}$, and apply this formula directly.  
Additional corrections due to flavour mixing are expected to 
be highly subleading.
We also neglect the contributions from neutralinos and
charginos, which are also expected to be less important~\cite{Heinemeyer:2004gx,
Heinemeyer:2007bw,Cho:2011rk}.

  A light stop can also contribute to flavour mixing, and to 
the branching fraction $\text{BR}({B}\to X_s\gamma)$ in particular.
This effect is present even if the scalar soft parameters
are all diagonal in the super-CKM basis (discussed in Appendix~\ref{sec:mfv})
due to loops involving a top and a charged Higgs or a stop and a chargino.  
These new contributions vanish in the supersymmetric limit, but can modify
the branching fraction when supersymmetry breaking is 
present~\cite{Barbieri:1993av,Okada:1993sx,Degrassi:2000qf}.
In our analysis, we use the HFAG experimental result
$\text{BR}(B\to X_s\gamma) = (3.55\pm 0.24\pm 0.09)\times 10^{-4}$~\cite{Amhis:2012bh}
together with the SM prediction of $\text{BR}(B\to X_s\gamma) 
= (3.15\pm0.23)\times 10^{-4}$ from Ref.~\cite{Grzadkowski:2008mf}.
We also consider limits from $\text{BR}(B_s\to \mu^+\mu^-)$~\cite{Aaij:2012ac} 
and $\Delta_0(B\to K^*\gamma)$~\cite{Hurth:2012jn,Amhis:2012bh}.

  We investigate the contributions of a light stop to 
$\text{BR}({B}\to X_s\gamma)$ and other flavour observables by
using SoftSUSY~3.3.5~\cite{Allanach:2001kg} to compute the superpartner
mass spectrum and mixings and passing the output to 
SuperISO~3.3~\cite{Mahmoudi:2008tp} to evaluate the modifications 
to flavour observables.  Input and output are achieved in the SLHA2
format~\cite{Allanach:2008qq}, and a detailed description of
the interface is given in Appendix~\ref{sec:mfv}.  

\begin{figure}[ttt]
${}$\vspace{-2cm}
\begin{center}     
\includegraphics[width=0.47\textwidth]{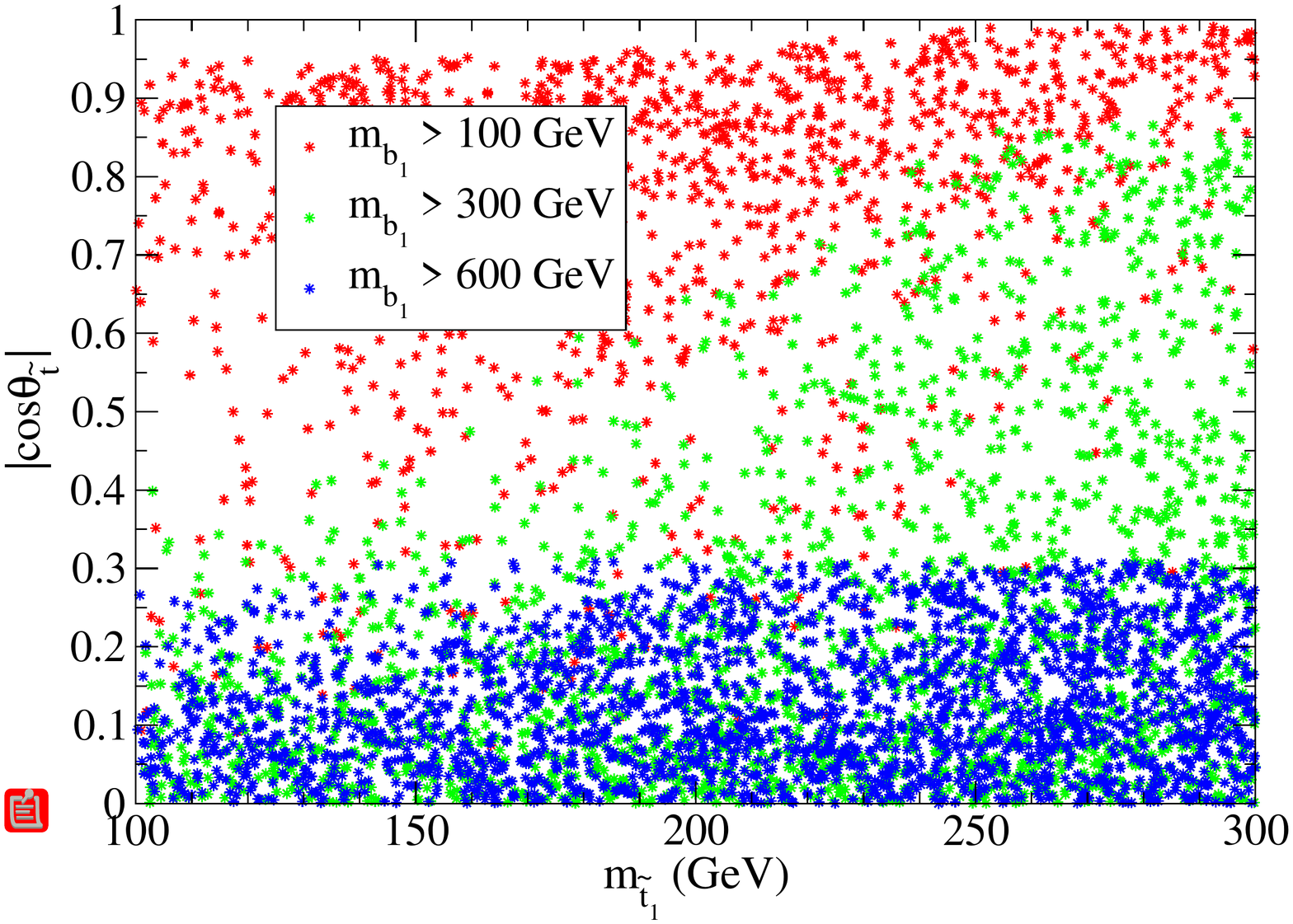} 
\phantom{A}
\includegraphics[width=0.47\textwidth]{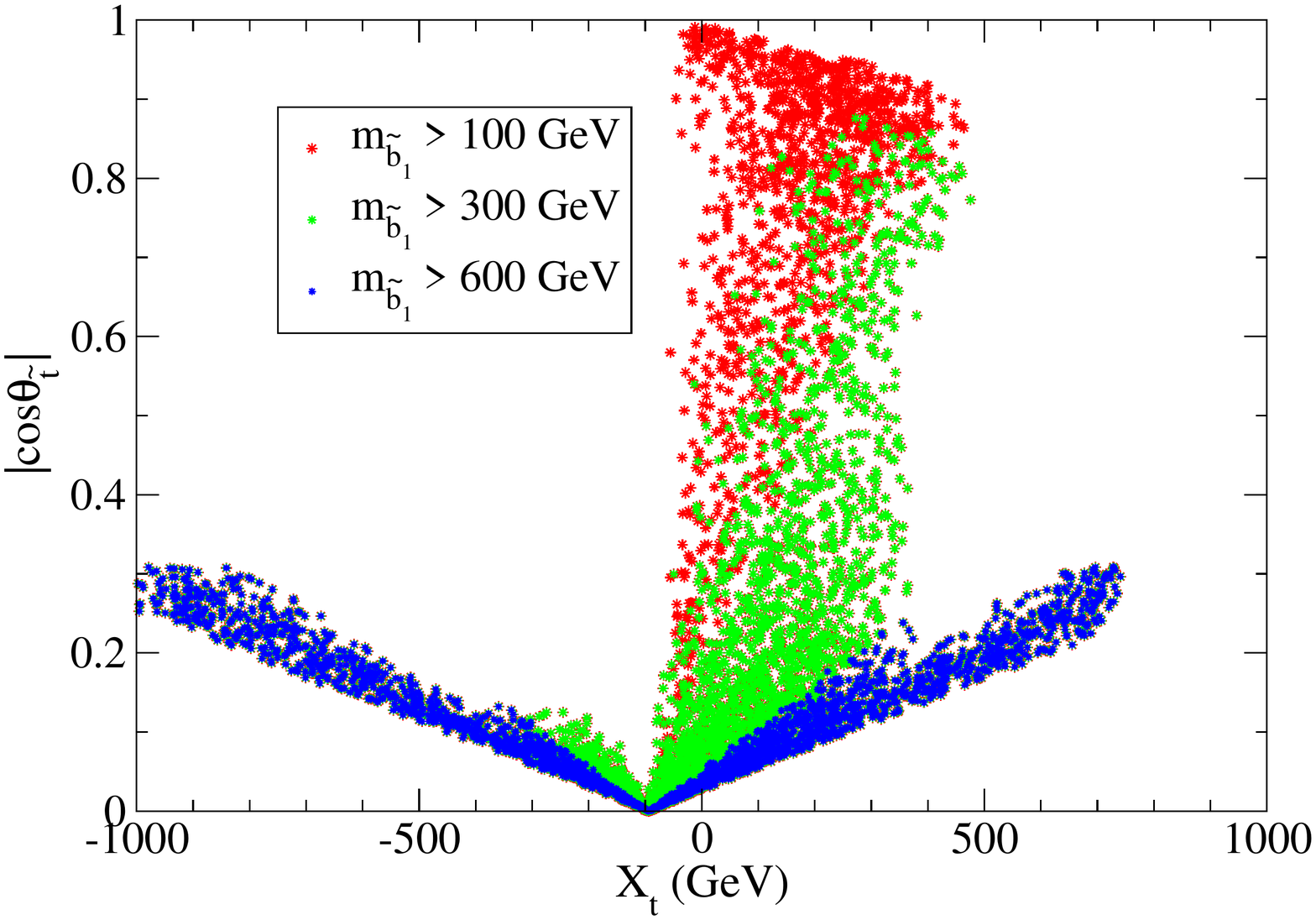}
\caption{
Allowed light stop mass and mixing angles for $\tan\beta=10$.
}
\label{fig:flpew-t10}    
\end{center}  
\end{figure}

  With these tools, we have performed a series of scans over stop 
parameters by varying the $A$, $b_i$, and $c_i$ MFV parameters,
while holding fixed $a_i=1$ and $\tilde{m}^2=(1500\,\gev)^2$ 
to maintain heavy first- and second-generation sfermions near $1500\,\gev$.
We have also set $M_3=1000\,\gev$, $M_1=200\,\gev$, $M_2 = 350\,\gev$,
and $\mu=M_A=500\,\gev$.  The resulting allowed regions, 
consistent with both precision electroweak and flavour constraints 
at the $2\sigma$-level,\footnote{The ranges are $\Delta\rho \in [-1.2,9.4]\times 10^{-4}$ and 
$\text{BR}(B\to X_s\gamma)\in [2.77,4.33]\times 10^{-4}$.} 
are shown in Fig.~\ref{fig:flpew-t10}
for different values of the stop mass $m_{\tilde{t}_1}$, the stop mixing
angle $|\cos\theta_{\tilde{t}}|$,\footnote{Our convention is 
$|c_{\tilde{t}}| = 1\;(0)$ when the lightest stop is purely
left(right)-handed.}
and the stop mass-matrix mixing parameter $X_t = X_{t_{33}}$.  
In both panels, we have fixed $\tan\beta = 10$ and we have restricted 
$100\,\gev < m_{\tilde{t}_1} < 300\,\gev$ and $m_{\tilde{b}_1}> 100\,\gev$.  
We also show the effects of imposing additional restrictions on the 
mass of the lightest sbottom: 
$m_{\tilde{b}_1} = 300\,\gev,\,600\,\gev$, which are 
motivated by the strong limits placed on sbottoms by 
the recent analysis of Ref.~\cite{Asbot}.

  Our primary conclusion from Fig.~\ref{fig:flpew-t10} is that 
electroweak and flavour bounds alone permit a broad range of mixing
angles for a light stop, but that demanding a heavier sbottom
forces the light stop to be mainly right-handed.\footnote{
A light sbottom might still be consistent with LHC searches if
it is highly degenerate with both the light stop 
and the neutralino LSP~\cite{Asbot}.}
Despite the broad range of allowed mixing angles, we also find 
a severe restriction on the stop mixing parameter $X_t$ for smaller 
values of $m_{Q_{33}}^2$, which comes partly from $\Delta\rho$ 
but mainly from $\text{BR}(B\to X_s\gamma)$.  We have also examined the
allowed ranges for $\tan\beta = 30$, and we obtain similar results,
albeit with a stronger restriction on $X_t$.
The limitations on $X_t$ derive from an interplay between the chargino-stop
and charged Higgs-top contributions to $\text{BR}(B\to X_s\gamma)$.  These must
cancel to some extent when there is a light stop in the spectrum,
with the specific condition depending mainly on $\mu$, $M_A$, $\tan\beta$,
and $X_t$.  Even so, at the expense of some amount of tuning between these
parameters, a broad range of stop properties is possible.

  Let us also mention that among the bounds from flavour mixing, 
we find that $\text{BR}(B\to X_s\gamma)$ 
gives the most significant limit for the parameters we studied,
and is more restrictive than $\text{BR}(B_s\to \mu^+\mu^-)$~\cite{Aaij:2012ac}
and $\Delta_0(B\to K^*\gamma)$~\cite{Amhis:2012bh}.
We also find that the contributions to flavour observables
are typically dominated by the CKM mixing already present in 
the super-CKM basis.  
More precisely, the deviations from flavour-diagonal 
squark mass matrices in the super-CKM basis allowed by MFV 
do not appreciably change the effects of the superpartners 
on the flavour observables we have studied.
Indeed, in the case of $\text{BR}({B}\to X_s\gamma)$, we find substantial 
agreement with SuSpect~2.41~\cite{Djouadi:2002ze} which does not include 
any additional (non-super-CKM) flavour mixing.

\subsection{Higgs Production and Decay Rates}

  Light stops are also closely linked to the mass of the SM-like
$h^0$ Higgs boson in the MSSM.  For many of the sets of stop masses we
consider in this work, the mass of $h^0$ lies well below
the mass of the tentative Higgs resonance near $125\,\gev$~\cite{
Hall:2011aa,Draper:2011aa}.
Within the MSSM, this discrepancy can be resolved by pushing 
the left-handed stop soft parameter $m_{Q_{33}}^2$ to values well 
above the TeV scale~\cite{Carena:2008vj,Carena:2012np}.  
This problem can also be addressed by expanding the field content to include
an additional singlet, as in the NMSSM.  In both cases, these fixes
can be achieved without necessarily altering the collider signatures 
of the light stop.  

  A light stop can also modify the production and decay rates of the
$h^0$ Higgs~\cite{Dawson:1996xz}.  When the light state is mainly unmixed,
the effect is to enhance the production rate in gluon fusion
and to suppress the branching fraction to photon pairs~\cite{
Menon:2009mz,Cohen:2012zza,Curtin:2012aa}.
Conversely, when the the light stop is highly mixed it has the opposite 
effect, pushing down the rate for gluon fusion and increasing
the branching ratio to diphotons~\cite{Dermisek:2007fi}.  The Higgs rates
can also be modified by the presence of other MSSM (or beyond) superpartners
that run in the loops~\cite{Carena:2012np,Carena:2012gp} 
or that mix with the Higgs~\cite{Ellwanger:2011aa,Batell:2011pz}.  
Again, these extensions of the MSSM need not modify the collider signals 
of the light stop.

  Light stops can also contribute in a more direct way to Higgs searches
by the formation of stoponium, a stop-antistop bound state~\cite{Drees:1993uw,Martin:2008sv,Barger:2011jt,Kats:2012ym}.  Stoponium decays can mimic those
of a Higgs boson.  However, for $m_{\tilde{t}_1} > m_{h^0}$, the dominant
stoponium decay mode is frequently to a pair of $h^0$ Higgs bosons
(particularly if the stop state is not highly mixed).  Current LHC searches
do not appear to be sensitive to this decay channel~\cite{Barger:2011jt}.
  
  For these various reasons, we do not constrain the spectrum of light stops
based on searches for and the tentative discovery of the $h^0$ Higgs boson.  
Instead, we focus on the direct LHC limits on a light stop state.

\section{Light Stop Decay Modes\label{sec:decay}}

We consider the decays of a light stop to the lightest neutralino 
or chargino in the spectrum, taking the rest of the superpartners to be heavy.
We also focus on the case where $\tilde{t}\to t\chi_1^0$ is kinematically
forbidden.  The stops then have four possible decay modes: 
\begin{enumerate}
\item two-body~(2B): $b \chi_{1}^{+}$.
\item three-body~(3B): $b W^+ \chi_{1}^{0} $.
\item four-body~(4B): $b f \bar{f}' \chi_{1}^{0}$. 
\item flavour-violating~(FV): $c \chi_{1}^{0}$. 
\end{enumerate}
Our analysis shows that each of these modes tends to dominate 
in a different kinematic region, depending on the stop-neutralino 
and stop-chargino mass differences.

\subsection{Flavour-Violating Stop Decays in MFV}

  The most model-dependent of these decays modes is the FV channel.
We apply the results of Ref.~\cite{Hiller:2008wp} to calculate the 
FV decay width within the framework of 
Minimal Flavour Violation~(MFV)~\cite{D'Ambrosio:2002ex}.
The general form of the stop-charm-neutralino coupling is~\cite{Hiller:2008wp}
\beq 
\mathscr{L}_{c \til t \til \chi_{1}^{0}} 
= \bar{c}(y_{L} P_{R} + y_{R} P_{L})\til \chi_{1}^{0} \til t + \rm h.c. \ , 
\eeq
in terms of which the FV decay width is given by
\beq 
\Gamma = \frac{ Y^{2}}{16 \pi}\,m_{\til t}\,
\left(1 - \frac{m_{\til \chi_{1}^{0}}^{2}}{m_{\til t}^{2}}\right)^{2} 
\ , \label{eq:FVwidth} 
\eeq
where $Y = \sqrt{|y_{L}|^{2}+|y_{R}|^{2}}$.
   

  The couplings $y_L$ and $y_R$ can be computed reliably in the 
mass-insertion approximation.  The relevant flavour mixings in this
case arise from the following off-diagonal elements 
of the soft masses~\cite{Hiller:2008wp}:
\bea 
(\til m_{Q_L}^{2})_{23} &=& \til m^{2} b_{2} (\yd \yd^{\dag})_{23} = \til m^{2} b_{2} \lambda_{b}^{2} V_{cb} V_{tb}^{*} \nonumber \\
     (\til m_{U_R}^{2})_{23} &=& \til m^{2} c_{1} (\yu^{\dag} \yd \yd^{\dag} \yu)_{23} = \til m^{2} c_{1} \lambda_{t} \lambda_{c} \lambda_{b}^{2} V_{cb} V_{tb}^{*} \\
     (a_{u})_{23} &=& A b_{7} (\yd\yd^{\dag}\yu)_{23} = A b_{7} \lambda_{t} \lambda_{b}^{2} V_{cb} V_{tb}^{*} \nonumber \\
     (a_{u})_{32} &=& A b_{7} (\yd\yd^{\dag}\yu)_{32} = A b_{7} \lambda_{c} \lambda_{b}^{2} V_{cb} V_{tb}^{*}. \nnmb
\eea  
In writing these expressions we have neglected terms proportional
to the charm Yukawa.  These contributions are negligible
unless the LR stop mixing is tuned to nearly zero.
With these off-diagonal soft terms, the stop-charm-neutralino
couplings are
\beq 
y_{L} &=& \frac{\lambda_{b}^{2} V_{cb} V_{tb}^{*}}{m_{\til t}^{2} 
- m_{\til c_{L}}^{2}}(\til m^{2} b_{2} 
c_{\til t} + A v_{u} b_{7} \lambda_{t} s_{\til t})
(-\frac{g'}{3\sqrt{2}} {N}_{11} - \frac{g}{\sqrt{2}} {N}_{21}), \label{eq:yL}\\
y_{R} &=& \mathcal{O}(\lambda_c) \label{eq:yR}
\eea 
where $c_{\til t}=\cos \theta_{\til t}$ and $s_{\til t}=\sin \theta_{\til t}$ 
correspond to the LR stop mixing, $\vu$ is the up-type Higgs expectation value, 
and ${N}_{ij}$ is the neutralino mixing matrix. 
Note that the relevant contributions to the FV decay width 
correspond to the stop mixing into the left-handed
charm squark $\tilde{c}_L$.
All the contributions involving the right-handed $\til c_R$ squark 
are proportional to the charm Yukawa, which are subleading 
and are neglected here.

  We compute the FV decay widths by calculating the mass spectrum
(without flavour mixing) with SuSpect~2.41~\cite{Djouadi:2002ze}.
Comparing with SoftSusy~3.3.5, corrections to the masses and
the LR mixing from off-diagonal soft terms are small provided the
flavour-diagonal terms match up.  Using these masses, we then apply
Eqs.~(\ref{eq:FVwidth},\ref{eq:yL},\ref{eq:yR}) to obtain the decay width.


\subsection{Other Decay Modes}

  For the 2B, 3B, and 4B modes, we use SuSpect 2.41~\cite{Djouadi:2002ze} 
to calculate the mass spectrum and we interface it with 
SUSY-HIT~\cite{Djouadi:2006bz} to find the 
decay widths.  While the 2B and 3B channels are treated in full, 
the SUSY-HIT implementation of the 4B mode is an approximation
to the full result~\cite{Boehm:1999tr} where the outgoing 
$f\bar{f}'$ fermion decay products are treated as massless.
We expect this to be a good approximation except for very close
to the kinematic boundary where we apply a phase space factor
to account for the fermion masses.

\subsection{Phase Diagrams}

  With the decay widths in hand, we can chart out the dominant decay
channels in different regions of the underlying parameter space.  
We find that the most likely decay channels are 
determined largely by phase-space considerations, which depend in turn
on the stop, chargino, and neutralino masses.
When the 2B and 3B channels are disallowed, we also find that
the 4B mode can overcome the FV mode within the MFV framework.

\begin{figure}[ttt]
${}$\vspace{-2cm}
\begin{center}     
${}$\hspace{-1.5cm}
\includegraphics[width=0.47\textwidth]{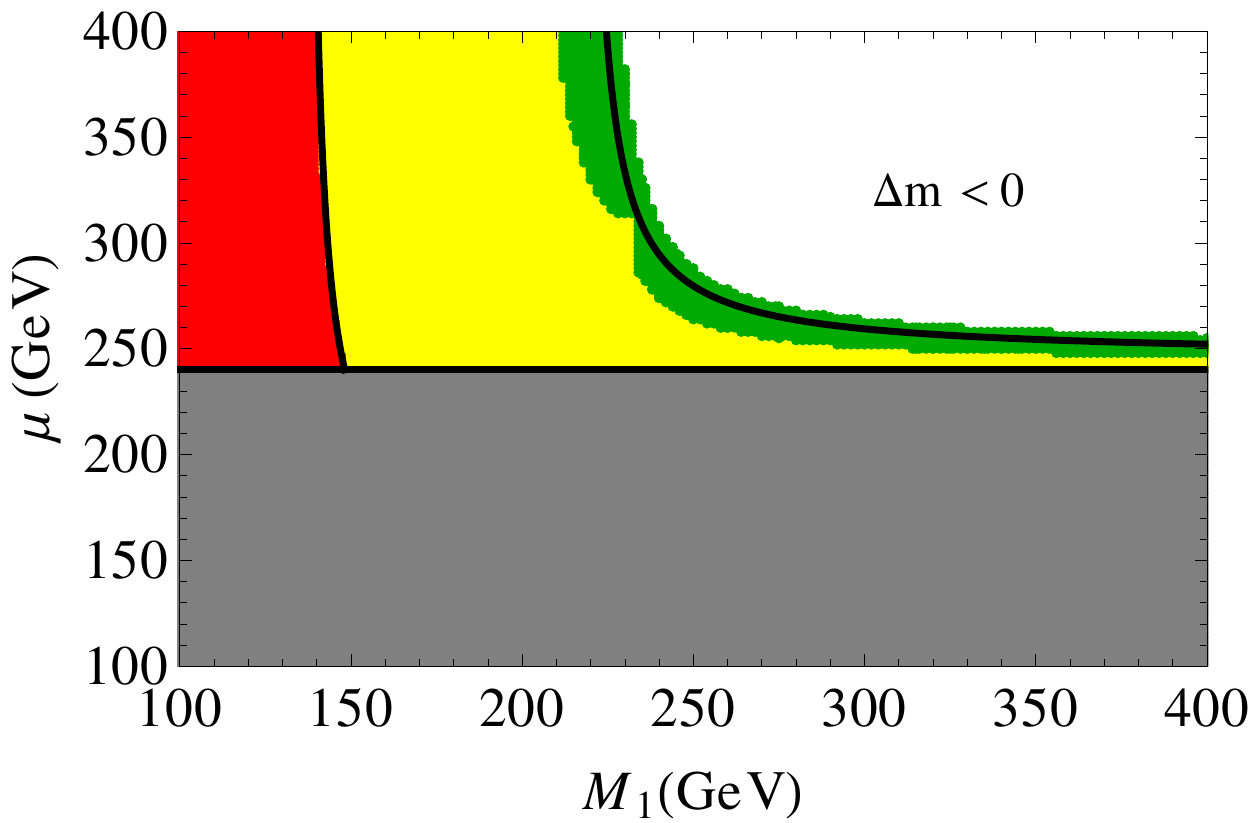} 
\phantom{A}
\includegraphics[width=0.47\textwidth]{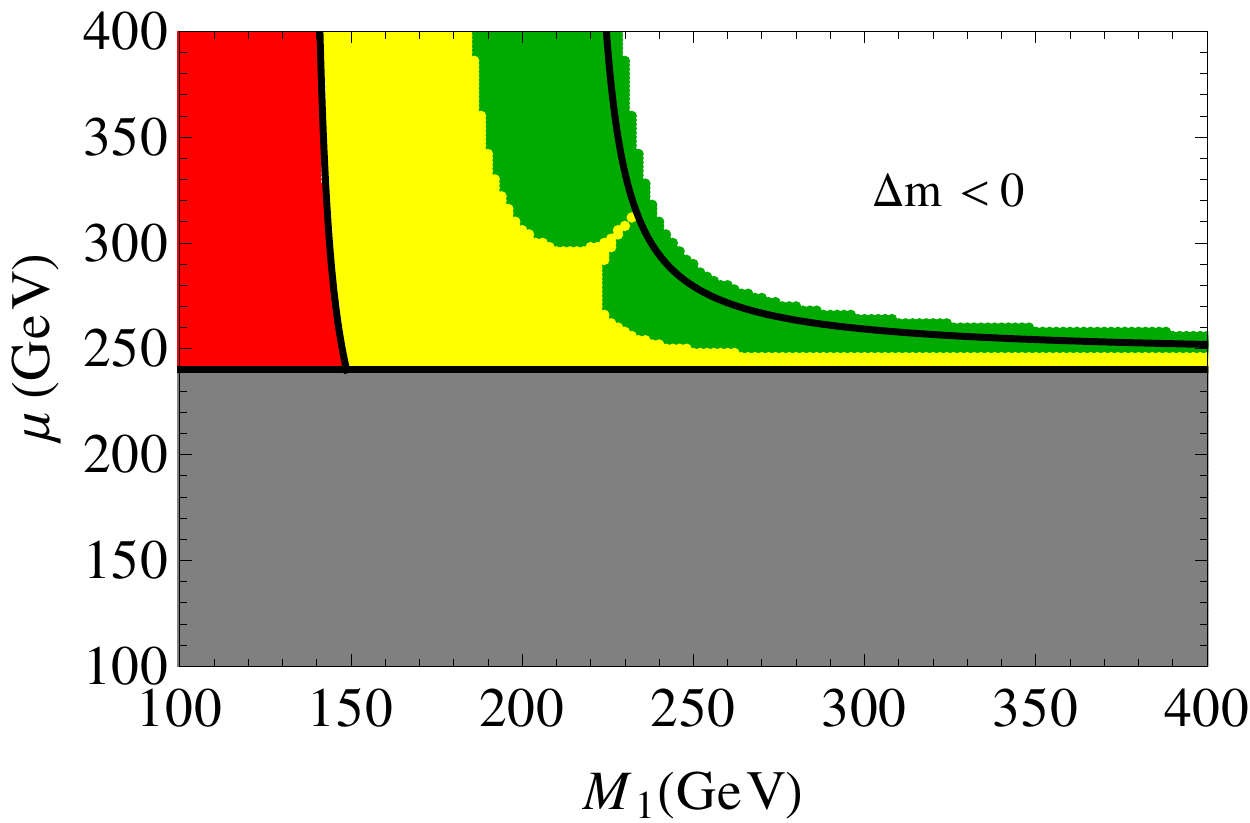}
\caption{
Dominant stop decay modes of a stop with mass
$m_{\tilde{t}} = 225\,\gev$ for $\tan\beta=10$, 
with $\cos\theta_{\til t} = 0.1$~(left) 
and $\cos\theta_{\til t} = 0.7$~(right).   
The four regions correspond to dominant decays by the 2B~(gray), 
3B~(red), 4B~(yellow) and FV~(green) channels.  We also show 
the kinematic boundaries between these decay channels by solid black lines.
}
\label{fig:phasediagrams-10}    
\end{center}  
\end{figure}

  In Fig.~\ref{fig:phasediagrams-10} we show the dominant decay channel 
of the lightest stop in the $M_1\!-\!\mu$ plane for a stop mass
of $m_{\tilde{t}} = 225\,\gev$, $\tan\beta=10$, and mixing angles of 
$\cos\theta_{\til t} = 0.1$~(left) and $\cos\theta_{\til t} = 0.7$~(right).
The four regions in the plots correspond to dominant decays by the 2B~(gray), 
3B~(red), 4B~(yellow) and FV~(green) channels.  The phase space
boundaries between the various decay channels are shown with solid black
lines.  In these plots we have also fixed $M_2 = 350\,\gev$, $M_A=500\,\gev$ 
and $b_2=b_7 = 0.5$.  All other input terms are as described 
at the end of Sec.~\ref{sec:stop}.  

  From these plots we see that the $2B$ channel dominates for smaller
values of $\mu$, when $m_{\tilde{t}} > m_b+m_{\chi_1^+}$ and the decay is open.  
This is to be expected, since the channel is neither suppressed 
by phase space nor flavour mixing.
As this mode turns off, the 3B channel comes to dominate at smaller
values of $M_1$ where it is allowed, up to $m_{\tilde{t}}> m_b+m_W+m_{\chi_1^0}$.
For larger $M_1$, the $3B$ channel shuts off as well and the $4B$ 
and $FV$ channels take over.  The boundaries between these regions 
are also nearly identical to the phase space boundaries.
Note that the $4B$ channel is cut off when 
$(m_{\til t}-m_{\chi_1^0}) < m_{b}$, in which case the FV mode is all
that is left.

  The interplay between the 4B and FV decay modes is more complicated.
While the 4B mode has significant suppression from the four-body final-state
phase space, the FV channel is reduced by the level of flavour mixing.
Very approximately, the relative suppression factors are 
$1/4!(4\pi)^4\sim 2 \times 10^{-6}$ for the 4B decay and
$(m_b\tan\beta/v)^4|V_{cb}|^2 \sim 5\times 10^{-6}(\tan\beta/10)^4$.
However, FV decays can receive further suppression if the light
stop is mostly right-handed, or they can be enhanced for larger
values of $\tan\beta$.  

  Comparing the left and right panels of Fig.~\ref{fig:phasediagrams-10}, 
the enhancement of the FV channel relative to 4B with increasing 
$\tilde{t}_L$ component  is evident.  When the lightest state
is mostly right-handed, as in the left panel of the figure where 
$\cos\theta_{\tilde{t}} = 0.1$, the 4B decay dominates
over the FV nearly up to the kinematic limit.  
For this reason, together with the mass hierarchy between the 
stop and the scharm, we obtain a larger branching fraction
in the 4B mode relative to previous 
analyses~\cite{Hikasa:1987db,Muhlleitner:2011ww}.

  When the stops are maximally mixed, $\cos\theta_{\tilde{t}}=0.7$,
the FV channel takes over well beyond the kinematic boundary.
This trend continues as the left-handed content of the light stop
is increased further.  However, this also gives rise to a light sbottom
squark and a new decay channel $\tilde{t}\to \tilde{b}W^{+(*)}$
that can dominate in what would otherwise be the 4B and 3B regions.
Given the very strong direct-search constraints on a light sbottom
as well as the indirect bounds from precision electroweak and flavour mixing,
we do not consider this possibility here.  

\begin{figure}[ttt]
\vspace{-2cm}
\begin{center}     
${}$\hspace{-1.2cm}
\includegraphics[width=0.47\textwidth]{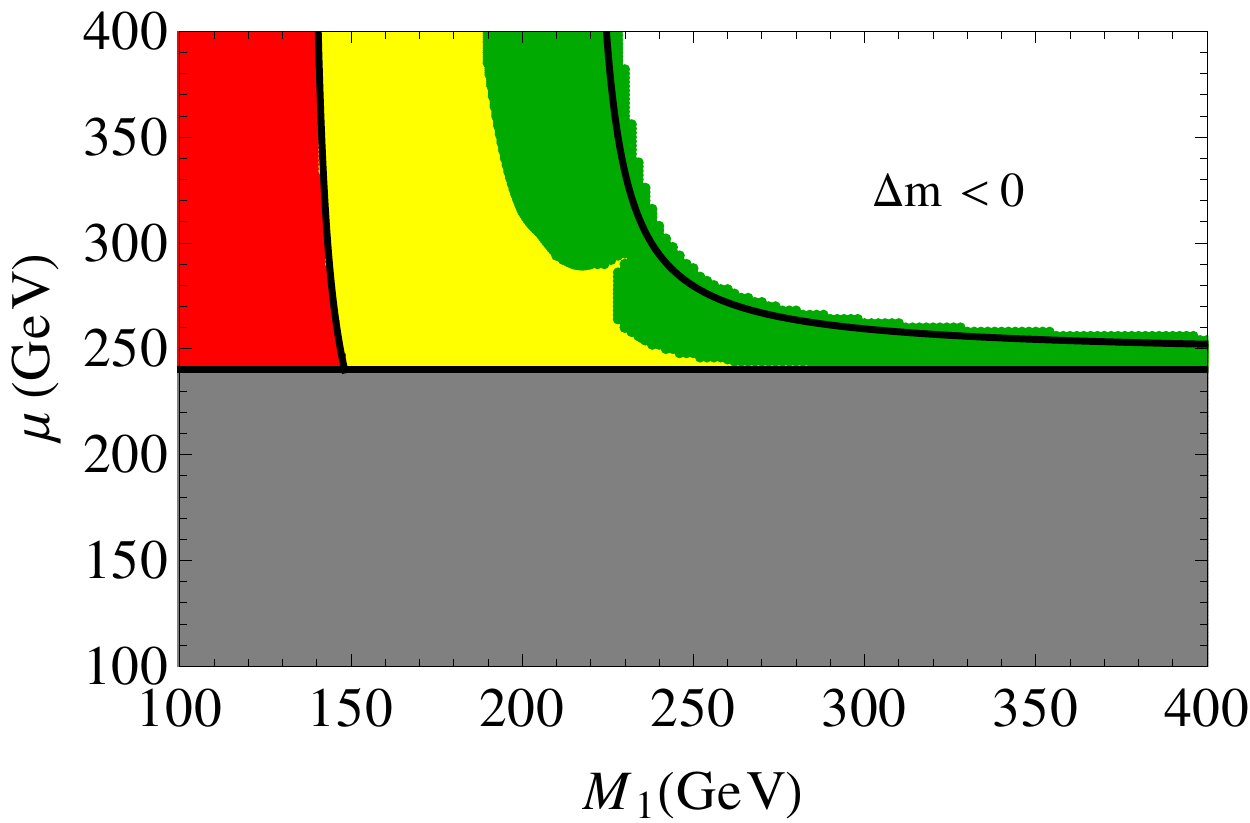} 
\phantom{A}
\includegraphics[width=0.47\textwidth]{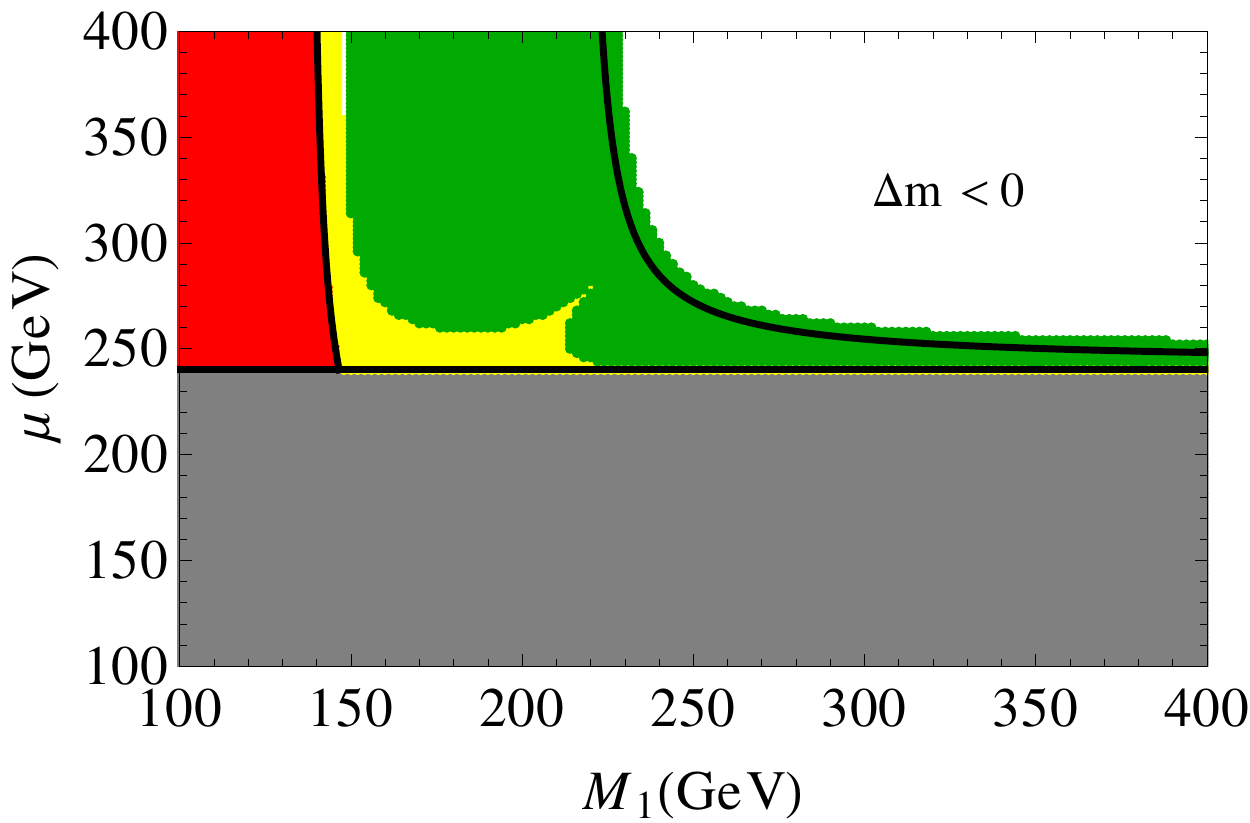}
\caption{
Dominant stop decay modes of a stop with mass
$m_{\tilde{t}} = 225\,\gev$ for $\tan\beta=30$, 
with $\cos\theta_{\til t} = 0.1$~(left) 
and $\cos\theta_{\til t} = 0.7$~(right).   
The four regions correspond to dominant decays by the 2B~(gray), 
3B~(red), 4B~(yellow) and FV~(green) channels.  We also show 
the kinematic boundaries between these decay channels by solid black lines.
}   
\label{fig:phasediagrams-30}    
\end{center}  
\end{figure} 

  The FV mode can also be enhanced relative to the 4B mode for larger
values of $\tan\beta$.  In Fig.~\ref{fig:phasediagrams-30} we show
the dominant stop decay channels for nearly the same parameters
as in Fig.~\ref{fig:phasediagrams-10}, $m_{\tilde{t}} = 225\,\gev$,
$M_2=350\,\gev$, $b_2=b_7 = 0.5$, but now with $\tan\beta=30$
(instead of $\tan\beta=10$).  In the left panel we have
$\cos\theta_{\til t} = 0.1$, and $\cos\theta_{\til t} = 0.7$ in the right.
As suggested by our simple estimates, the FV channel dominates over
a significantly larger region than before.  We conclude that, in MFV, 
either the FV or 4B decay modes can dominate in this region of the
superpartner spectrum, and it is important to search for both.  

  The absolute decay widths obtained in the FV-4B region receive
a significant amount of suppression.  Even so, we find that they
are generally prompt on collider time scales unless the stop-LSP
mass difference becomes very small, in agreement with Ref.~\cite{Hiller:2008wp}.
For our collider studies, we will assume that all the decays are prompt.
However, the emergence of a displaced stop decay vertex could allow 
for the direct measurement of the underlying flavour 
structure~\cite{Hiller:2008wp}, and would be interesting to study in its
own right.  On the other hand, a stop that is long-lived on the time scale
of the LHC detectors is firmly ruled out to masses well over 
$m_{\tilde{t}_1}> 500\,\gev$~\cite{Chatrchyan:2012sp}.

\section{LHC Search Sensitivity\label{sec:lhc}}

  In this section we investigate the sensitivity of existing LHC searches
to a very light stop decaying via the FV, 4B, or 3B modes.  
We make use of public ATLAS and CMS analyses to estimate  
exclusion limits in the $\mstop-\mneut$ plane for stop masses 
in the range 100-350 GeV.  Although these experimental studies
were not optimized for light stops, we find that they imply quite 
stringent bounds on such a state. 

  Throughout our investigation, we treat the light stop within a simplified
model where it decays exclusively (and promptly) via the FV, 4B, or 3B 
modes to a neutralino LSP and SM particles within the corresponding 
kinematically relevant region of the $\mstop-\mneut$ plane.  
As seen in the previous section, 
varying the LR stop mixing, $\tan\beta$, or the amount of flavour mixing
can modify the relative branching ratios of the FV and 4B channels.\footnote{
While these decay modes are usually prompt, they are slow enough that
the stop will hadronize before it decays.  We do not attempt to account
for this effect, but we expect that it will only be significant close
to the kinematic cutoffs based on the typical momenta of the decay products.}
Exclusion limits for intermediate cases can be obtained by interpolating 
between the separate FV and 4B cases studied here.
We do not consider light sbottoms, or production via gluinos 
or other superpartners.

\subsection{Event Generation and Simulation}

  We generate $pp\to \tilde{t}\tilde{t}^*$ events at $\sqrt{s}=7~\tev$ 
using MadGraph5 v1.4.3~\cite{Alwall:2011uj}.  The parton-level
output is passed to Pythia~6.4~\cite{Sjostrand:2006za} for decays, showering, 
hadronization, and matching of parton showers to matrix elements. 
For matched samples, the MLM scheme is used with shower-$k_{\perp}$ 
and $Q_{\rm match}=50~\gev$.   

  Simulation of the LHC detectors is performed with Delphes~\cite{Ovyn:2009tx}, 
using the anti-$k_{T}$ jet reconstruction algorithm.  
ATLAS and CMS detector/trigger cards are applied to obtain a more 
realistic comparison of our analyses to the respective LHC searches. 
We have cross-checked our results with PGS4 (using cone jet reconstruction 
with $\Delta$~R=0.7)~\cite{pgs}. The differences in the distributions 
of hardest jet $p_{T}$ and missing energy~($\MET$) are within
generally within 5-10$\%$ for the two detector simulators.  
In terms of acceptances to various searches, Delphes tends to 
be somewhat lower than PGS.

  In generating stop samples for our analysis, we are faced with a practical
challenge.  Many of the existing LHC SUSY searches are geared towards
finding relatively heavy coloured states that decay to a much lighter LSP.
These searches frequently require multiple hard jets and apply strong
cuts on $\MET$ or $m_{eff}=\Sigma_{j}p_{T}+\MET$.  As such, they
are typically only sensitive to the FV or 4B stop decay modes in certain
specific regions of the stop-production phase space.

  Since we are investigating relatively light stops, the typical energy
scale in the events tends to be somewhat small.  Furthermore, the momenta
of the stop decay products are distributed between multiple final states
in the 4B channel, and are also limited in the FV channel when the
stop-LSP mass difference is small.  Thus, the light stop events 
usually produce soft decay products and small amounts of $\MET$, 
particularly if the LSPs are emitted back-to-back.  
For these light stop events to pass many of the LHC SUSY analysis cuts, 
they must be produced in conjunction with one or more hard jets.
In addition to increasing the total energy in the event, 
the radiated jet gives a net boost to the $\til t\til t^{*}$ system
which can lead to an observationally significant amount of missing 
energy~\cite{Alwall:2008ve,Alwall:2008va}.

  The challenge, therefore, is to simulate efficiently this small corner
of the stop production phase space where one or more hard jets 
are emitted without being swamped by the much larger portion where 
the additional jets are relatively soft.
For this, we investigated the following generator-level samples:
\beq 
\til t\til t^{*}, \qquad \til t\til t^{*}+1j\,({\rm unmatched}), 
\qquad \til t\til t^{*}+1j\,(\rm{matched}), 
\nonumber
\eeq 
with matching scale $Q_{\rm match}=50~\gev$ for $1j$ matched,
and requiring $p_{T} > 50,100,150,200,$ $250,300~\gev$ for 
the hardest jet at the parton level in the unmatched samples.  
Of these, the matched sample is expected to best reproduce
the full $p_T$ distribution of the additional jet~\cite{Alwall:2008qv}.  
However, each given sample is strongly dominated by events where the extra
jet lies near the matching scale, leading to poor statistics
in the higher-$p_T$ region of greatest interest.  

  To improve the statistics of the samples we generate, we compared
the matched sample to a set of unmatched $\til t\til t^{*}+1j$ samples
constrained to have a jet $p_T$ greater than $50-300\,\gev$.
For our comparison, we studied the relative signals predicted by these samples
for a range of fiducial kinematic cuts typical of the 
ATLAS and CMS analyses considered in this paper: 
\begin{alignat}{2}
 \MET &> 130,\,200,\,300\,\gev  \qquad 
&\quad m_{eff} &> 500, 700, 1000\,\gev \nonumber \\  
 H_{T} &> 350, 500, 600\,\gev \qquad & p_{T,j1} &> 80, 130\,\gev \nonumber
\ . 
\end{alignat}
We find that the $\til t\til t^{*}+1j$ unmatched sample 
with $p_{T,j_1}>150~\gev$ provides a good balance between obtaining
useful event statistics (\emph{i.e.} a reasonable fraction of the generated
events pass the kinematic cuts) and not artificially eliminating the
signals.  For example, the unmatched sample with $p_{T,j_1}>100~\gev$
leads to the same effective signals to within $10\!-\!20\%$.

  We have also compared the $p_T$ distribution of hardest-jet in 
the $1j$ unmatched sample with $p_{T,j_1}>150~\gev$ to matched samples
with one or two additional hard jets.  Relative to the matched
$1j$ sample with $Q_{cut}=50\,\gev$, we find excellent agreement 
for $p_{T,j_1}>150~\gev$ (as is expected).  Comparing to the
matched $2j$ sample (matched with $H_T>100\,\gev$),
we also get excellent agreement for the $p_{T,j_1}$ distributions.
While the $2j$ matched sample tends to produce a harder second jet,
we find that including it does not appreciably increase the effective
signals that pass our sample kinematic cuts. 

  The stop production sample we generate arises at NLO and represents only
a small subset of the total stop production phase space.  As a result, 
we do not attempt to apply a K-factor to the cross section obtained
by comparing the inclusive LO and NLO cross sections (which is about
$1.4$ in this case based on Prospino2.1~\cite{Beenakker:1996ed}).
Instead, we estimate the effect of the theory uncertainty in the cross section
on the LHC exclusions by simply varying the MadGraph production cross
sections by $\pm 50\%$.

\subsection{LHC Searches and Exclusions}

  The LHC analyses that we apply to our light stop samples
are summarized in Tables~\ref{table:ATLASsearches} and 
\ref{table:CMSsearches}.  
These searches constrain different 
regions of the $\mstop-\mneut$ plane, and taken together,
provide significant bounds on a light stop.
All the searches considered have $\sqrt{s}=7\,\tev$.
We defer an analysis of the recent SUSY search results
obtained at $\sqrt{s}=8\,\tev$, which appeared as this paper
was nearing completion, to a future study.  

  We consider stop decays in the FV, 4B, and 3B modes, and the limits
from each LHC analysis are applied separately.  
In every case, we impose the selection requirements described
in the corresponding report to our simulated signal and we compare
to the listed backgrounds (accounting for the stated background uncertainties)
to obtain limits on the stop and neutralino masses.  
The corresponding 95\,\%\,c.l. exclusions are summarized 
in Fig.~\ref{fig:FV4Bdecays}~(FV, 4B) and Fig. \ref{fig:3Bdecays} (3B).
In these figures, we show the result for the central value of
the MadGraph-generated cross section with solid lines, and the 
exclusions when this cross section is varied by $\pm 50\,\%$
with dashed and dotted lines.  

  Note that many of the searches listed in Tables~\ref{table:ATLASsearches} 
and \ref{table:CMSsearches} do not yield a constraint.
This fact is shown in the rightmost column of the Tables,
where we list the decay channels for which the corresponding
searches are relevant; when no channel is listed, no interesting
bound was obtained.  In the discussion to follow, we will focus 
on the LHC analyses that do lead to limits.

\begin{table}[ttt]
\centering
\begin{tabular}{|c|c|c|}
\hline
ATLAS Analysis ($\mathcal{L}/\ifb$) & Final states & Decay channel  \\\hline\hline
Monojet (1.00) ~\cite{A1} & $\leq1$-$3j,0l$ & FV, 4B  \\ \hline
Jets+$\MET$ (1.04) ~\cite{A2} & $\geq 2$-$4j,0l$  & FV, 4B \\\hline
Monojet (4.7) ~\cite{A11}&  $\leq 2j,0l$  & FV, 4B     \\ \hline
$j+b+\MET$ (0.83) ~\cite{A3}  &  $\geq 3j,\geq 1b,0l$  &  -  \\\hline
$j+l+\MET$ (1.04) ~\cite{ATLAS:2011ad}  & $\geq$3-4$j,1l$  &  -  \\\hline
$j+b+l+\MET$ (1.03) ~\cite{A5} &  $\geq 4j$, $\geq 1b,1l$  &  -  \\\hline
$j+b+l+\MET$ (2.05) ~\cite{ATLAS:2012ah}  &  $\geq3$-$4j \geq 1b,0$-$1l$  &  -  \\\hline
$\til t \til t^{*} \rightarrow ll$ (4.7) ~\cite{A7}  &  $\geq 1j,1l$ & - \\\hline
$j+l+b+\MET$ (4.7) ~\cite{A12} &  $\geq 4j,1b,1l$  &  -  \\  \hline
$j+b+\MET$ (4.7) ~\cite{A13}  &  $\geq 4j,\geq 2b,1l$  &  -  \\ \hline
\end{tabular}
\caption{Summary of ATLAS searches at $\sqrt{s}=7 \tev$ that 
constrain the FV, 4B and 3B decay modes.}
\label{table:ATLASsearches}
\end{table}

\begin{table}[ttt]
\centering
\begin{tabular}{|c|c|c|}
\hline
CMS Analysis ($\mathcal{L}/\ifb$)  & Final states & Decay channel \\\hline\hline
Monojet (1.1) ~\cite{C1}   &  $\leq 2j,0l$  &  FV, 4B  \\ \hline
$j+\MET$ (1.1) ~\cite{C2}  &  $\geq 3j,0l$  &  -  \\ \hline
$\alpha_{T}$ (1.14) ~\cite{Chatrchyan:2011zy}  & $\geq 2j, 0$-$1l(\mu),0$-$1\gamma$ & - \\ \hline
$j+b+\MET$ (1.1) ~\cite{C4}   & $\geq3$-$4j,\geq 1b 1l, \geq 1b$ & FV, 4B \\ \hline
Razor (4.4) ~\cite{C5}     & $\geq 2j,0$-$2l$  & FV, 4B, 3B \\ \hline
$j+l+\MET$ (1.1) ~\cite{C6}  & $\geq 3$-$4j, 1l$  &   -  \\   \hline
$M_{\rm T2}$ (1.1) ~\cite{C7}     & $\geq 3$-$4j, \geq 0$-$1b, 0l$  &   FV, 4B, 3B  \\\hline
Monojet (5.0) ~\cite{Chatrchyan:2012me}   &  $\leq 2j, 0l$  & FV, 4B \\\hline
$j+\MET$ (4.98) ~\cite{:2012mfa}   &  $\geq 3j, 0l$  &  FV, 4B     \\\hline
$\alpha_{T}$ (4.98) ~\cite{C13}  & $\geq 2j, 0$-$3b, 0l,$ & -  \\\hline
$j+b+\MET$ (4.98) ~\cite{C14}  & $\geq 3j,1$-$3b, 0l $  & FV, 4B, 3B   \\ \hline
Razor (4.7) ~\cite{C15}     & $\geq 2j, \geq 1b,0$-$2l$  &   FV, 4B, 3B  \\ \hline
$M_{\rm T2}$ (4.73) ~\cite{:2012jx}     & $\geq 3$-$4j,\geq 0$-$1b,0l$  &  FV, 4B, 3B  \\\hline
\end{tabular}
\caption{Summary of CMS searches at $\sqrt{s}=7 \tev$ 
that constrain the FV, 4B and 3B decay modes.}
\label{table:CMSsearches}
\end{table}

\subsubsection{FV mode}

  The FV mode has $\tilde{t}\to c\chi_1^0$, which suggests that 
it will be probed most effectively by searches for jets and missing energy.
Thus, we apply the ATLAS and CMS monojet, jets+$\MET$, 
Razor, and $M_{T2}$ analyses to it.  However, due to a sizable mistag
rate for $c$ jets to be identified as $b$ jets (about $10\,\%$),
we find that the FV mode is also constrained by several $b$-jet analyses. 
In contrast, we find that searches with leptonic final states are
generally not sensitive to this decay channel.
The regions of the stop-neutralino mass plane excluded by the analyses
we consider for the FV decay mode are shown in the left set of
panels in Fig.~\ref{fig:FV4Bdecays}.

  Monojet searches require at least one hard jet and missing energy, 
and they are good probes of the small $\mstop-\mneut$ region where 
the outgoing charm quark is soft and difficult to detect~\cite{Carena:2008mj}.
In this case, the hard jet arises from QCD radiation.
ATLAS and CMS monojet searches apply various sets of cuts on the transverse
momentum of the hardest jet $p_{T\,j_1}$ and the missing energy $\MET$.
They also veto on events with hard leptons, as well as events with 
too many additional hard jets~\cite{A1,A11,C1,Chatrchyan:2012me}.  
We find that the harder cuts on $p_{T\,j_1}$ and $\MET$ used by these analyses,
which range up to $500\,\gev$, tend to be the most effective at probing 
the stop signal.  Among the set of events that pass the 
selection requirements, the stops must be created with a very hard extra
jet.  This can give the stop pair a large $p_T$, which leads
to significant missing energy but can also make the charm jets visible. 
From this point of view, the less-stringent veto on additional jet activity 
in Refs.~\cite{A11,Chatrchyan:2012me} relative to Ref.~\cite{A1} 
is beneficial.\footnote{The bounds from the monojet searches are also
likely to be valid when the stop decay is slightly displaced.}

  The presence of one or more hard QCD jets and the resulting boost of
the stops can lead to multiple hard jets in FV stop events.  Thus,
searches for multiple jets and missing energy can also be sensitive
to light stops.  The recent ATLAS jets+$\MET$ analysis of
Ref.~\cite{A2} divides their search into $2\!-\!4\,j$ channels,
and imposes a lepton veto and cuts on $\MET$ and 
$m_{eff} = \sum_{j} p_{T_j} + \MET$.
The $m_{eff}$ and $\MET$ requirements of this analysis are geared
towards heavier superpartners and are quite severe, with $m_{eff}$
typically greater than $1000\,\gev$, limiting their sensitivity.
In fact, the more recent ATLAS analysis of Ref.~\cite{A12} 
with $4.7\,\ifb$ of data imposes even stronger
requirements on $m_{eff}$ and is less sensitive than Ref.~\cite{A2}, despite
having much more data.
The kinematic requirements for the corresponding CMS searches for jets+$\MET$
are less severe, and give slightly stronger exclusions.
The CMS search for jets and missing energy of Ref.~\cite{:2012mfa}
requires at least 3 jets, $H_{T} = \sum_{j} p_{T_j} > 500~\gev$, 
missing energy, and no hard leptons.  This search places strong 
limits on cross-sections for new physics, excluding a region of the 
$\mstop-\mneut$ plane that is complementary to the monojet analyses.

  The FV mode is also constrained by searches involving $b$ jets
due to the significant ($\sim 10\%$) mistag rate of $c$-jets. 
The minimal requirements of the CMS jets+b+$\MET$ search~\cite{C4},
at least three jets with $p_T>50\,\gev$ and a $b$ tag, 
can be satisfied in this case.\footnote{
A small fraction (about $5\%$) of the stop events passing the cuts 
have genuine $b$ jets produced in the parton shower.}
This analysis also has weaker cuts on $H_T$ and missing energy
(limited via the $\alpha_T$ variable) relative to the jets+$\MET$ searches,
which leads to a significant exclusion of the FV mode.
Let us also point out that the observation of a light squark in 
multiple light- and heavy-quark flavour channels could provide
information about the flavour structure of the stop decay.
Even though the $c$ quark is misidentified in this case, 
the decay could still be differentiated from other flavour-violating
channels such as $\til t \rightarrow u\chi_1^0$, 
(which is negligible in MFV), 
since the mistag rate of $u$-jets as $b$-jets is about $1\%$.

  Light stops are also constrained by the CMS Razor and $M_{T2}$ searches.
The Razor variables provide a way to use event-shape information
to reduce backgrounds while not imposing overly strict kinematic
requirements~\cite{C5,C15}.  
The stop signal populates mainly the $M_R<1000\,\gev$ region,  
and the analysis of Ref.~\cite{C5} restricts a significant part
of the stop-LSP mass plane.  
  The CMS $M_{T2}$ analyses of Refs.~\cite{C7,:2012jx} have two separate 
signal regions, with and without a $b$-tag requirement.
Applying the analysis of Ref.~\cite{C7}, the $\emph High M_{T2}$ search channel 
($\geq 3j, 0b$) excludes a greater region than the 
$\emph Low M_{T2}$ channel ($\geq 4j,\geq 1 b$) 
due to the $b$-tag requirement of the latter. 
Similar exclusions are found for the more recent search 
of Ref.~\cite{:2012jx}.  While this analysis has considerably
more data than Ref.~\cite{C7}, it also imposes more stringent
kinematic requirements leading to about the same exclusion.

  Our results for the FV mode appear to be compatible with
previous analyses of this channel in the monojet and jets+$\MET$
channels~\cite{Bi:2011ha,Ajaib:2011hs,He:2011tp,Yu:2012kj,Dreiner:2012sh},
although Ref.~\cite{Choudhury:2012kn} obtains a somewhat stronger limit
from the jets+$\MET$ channel than we do.  The limits obtained in 
the MT2/Razor analyses also appear to be consistent with 
Ref.~\cite{Dreiner:2012sh}, while our jets+b+$\MET$ search is new
and provides a slightly stronger exclusion.

\begin{figure}[ttt]
${}$\vspace{-3cm}\\
\begin{center}    
\includegraphics[width=0.47\textwidth]{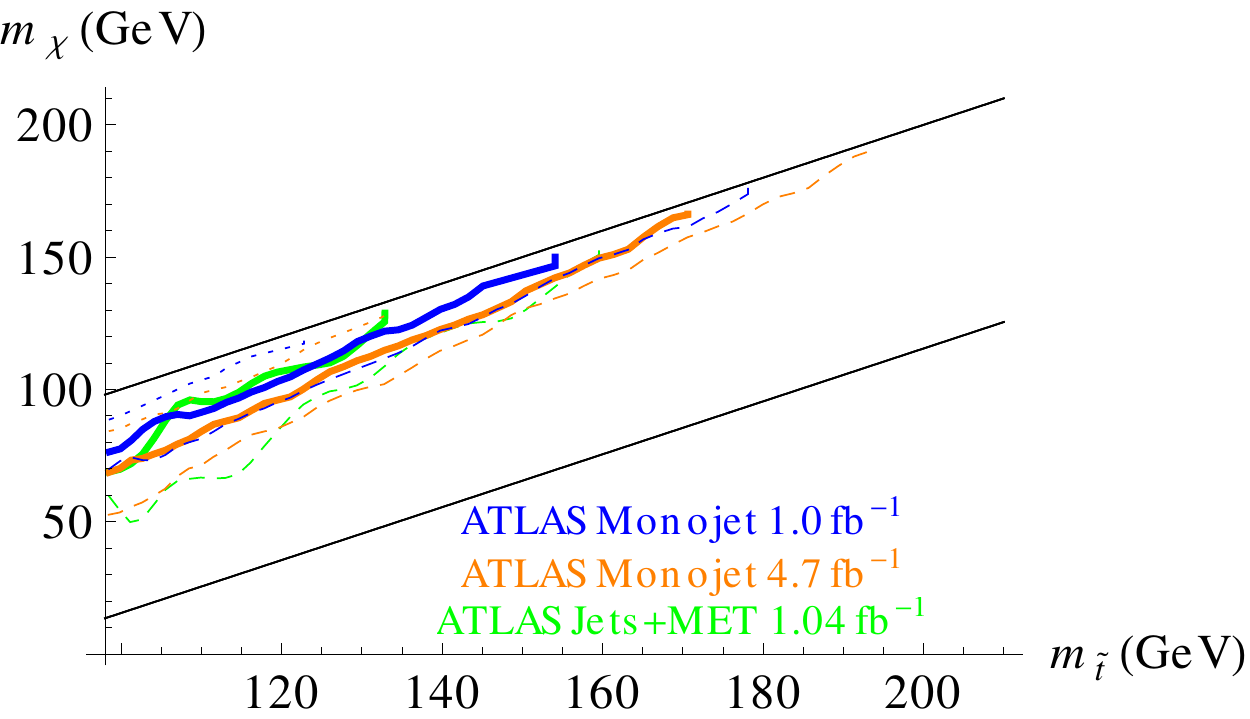}
\includegraphics[width=0.47\textwidth]{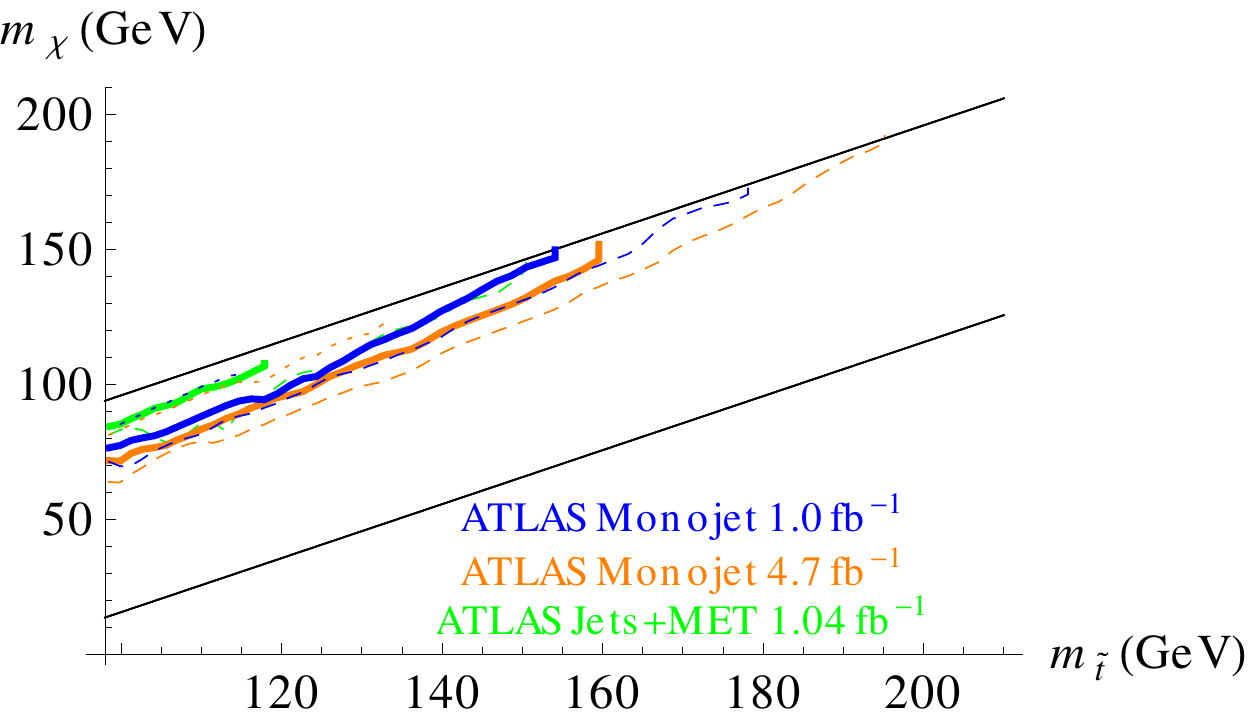}
\includegraphics[width=0.47\textwidth]{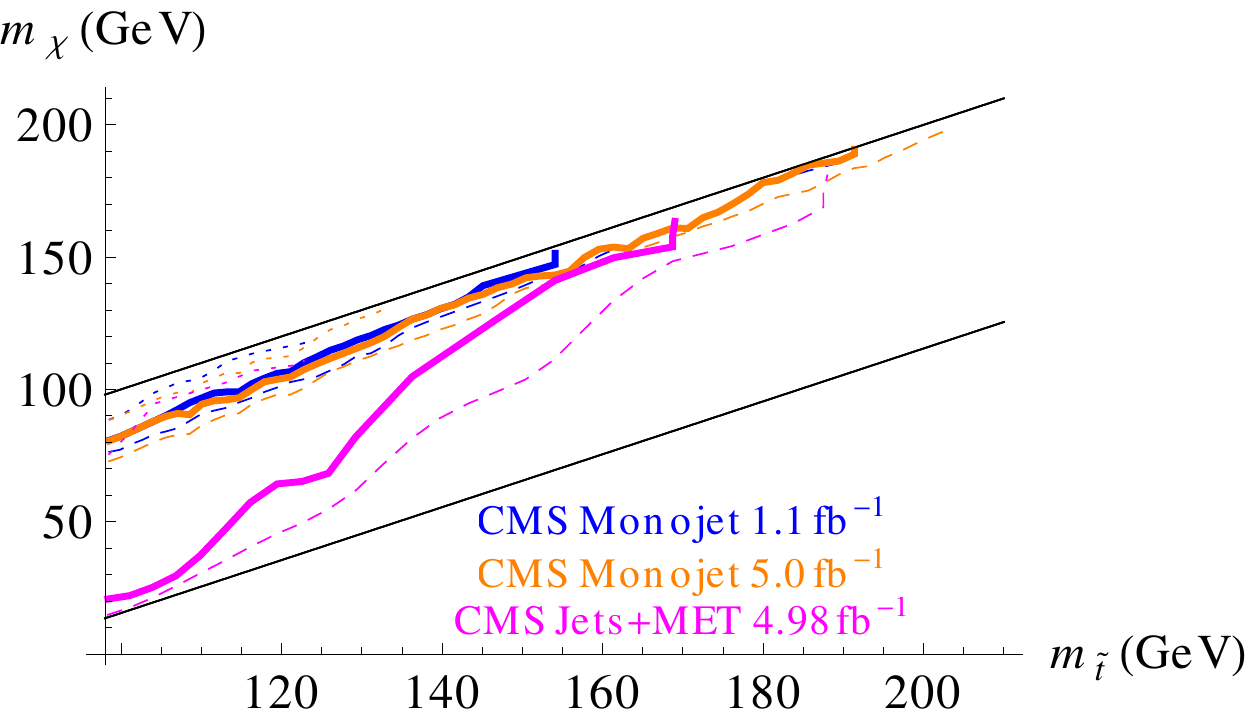}
\includegraphics[width=0.47\textwidth]{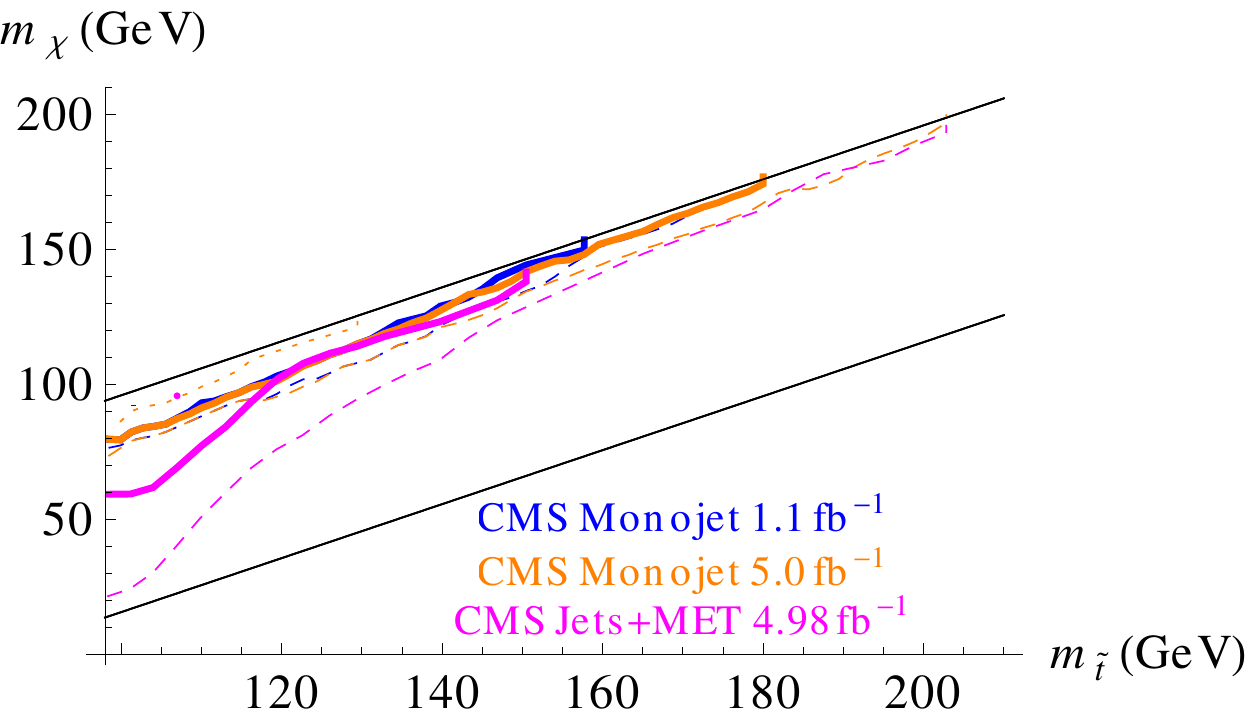}
\includegraphics[width=0.47\textwidth]{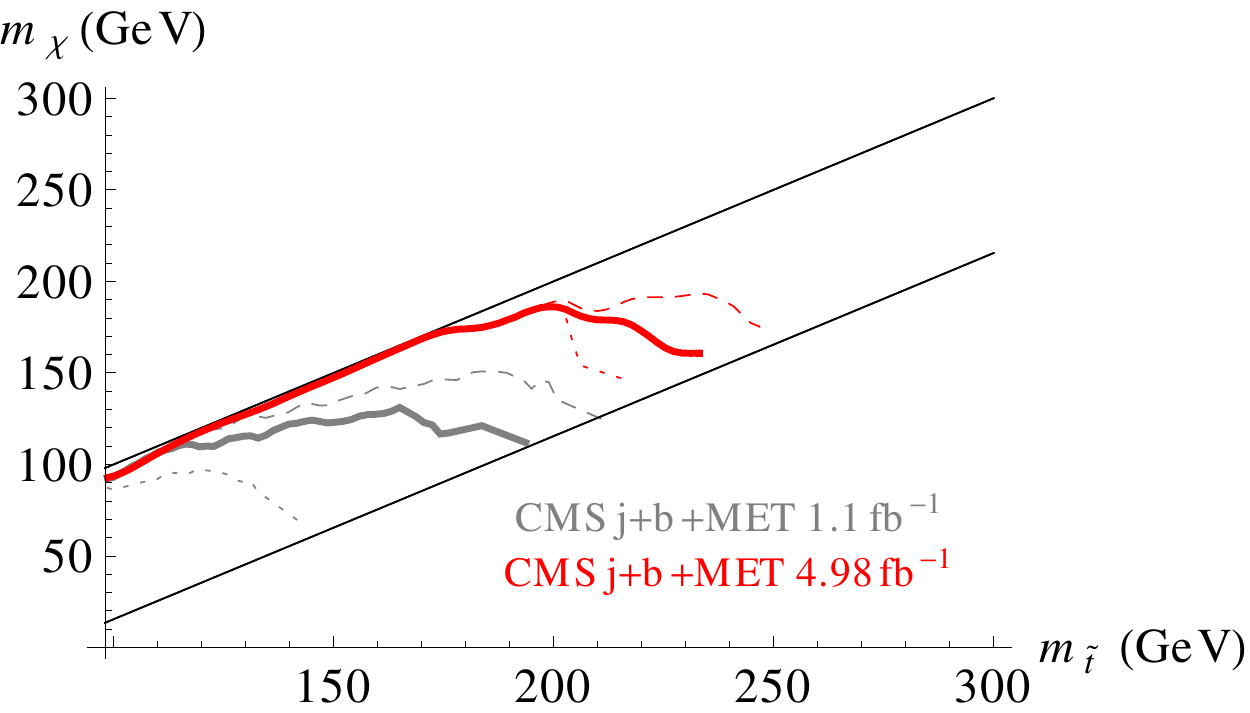}
\includegraphics[width=0.47\textwidth]{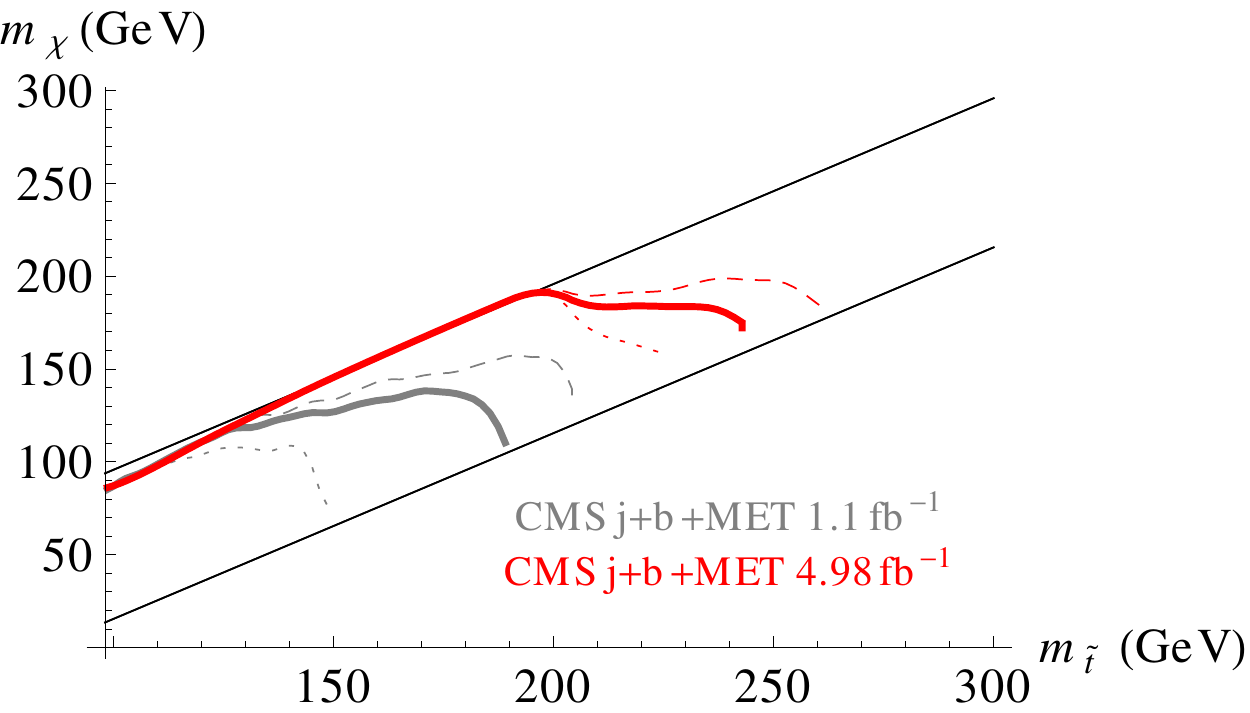}
\includegraphics[width=0.47\textwidth]{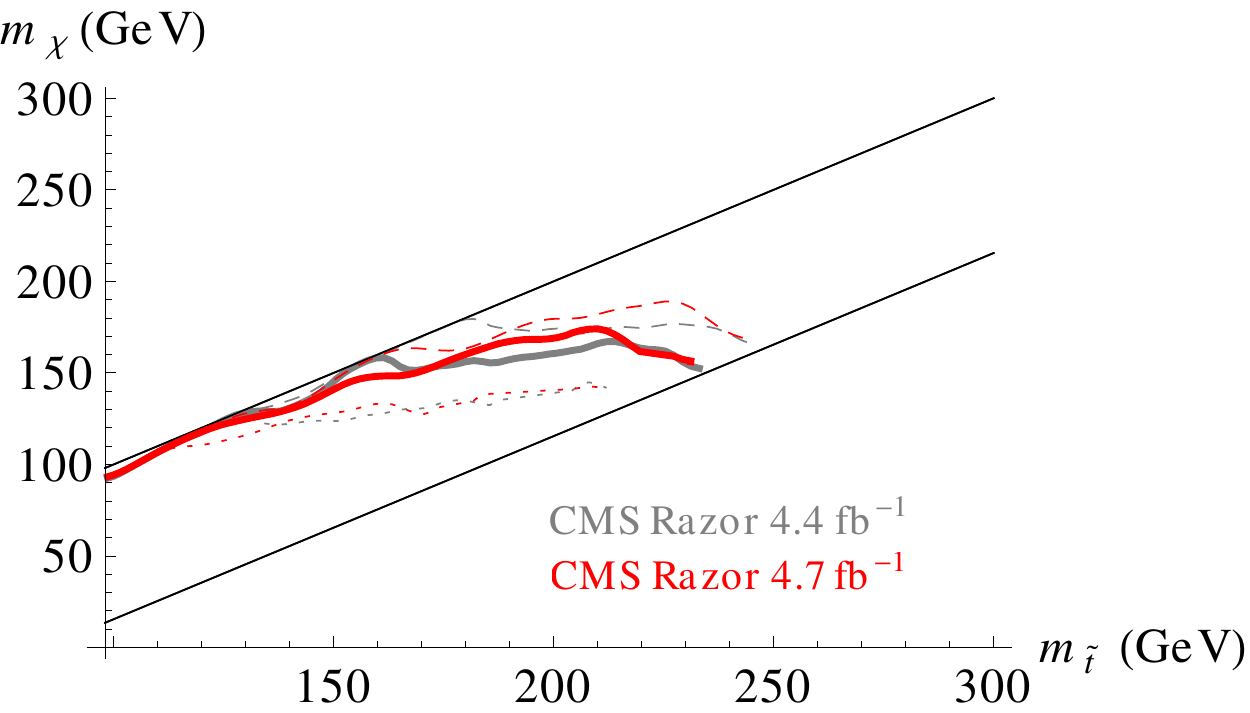}
\includegraphics[width=0.47\textwidth]{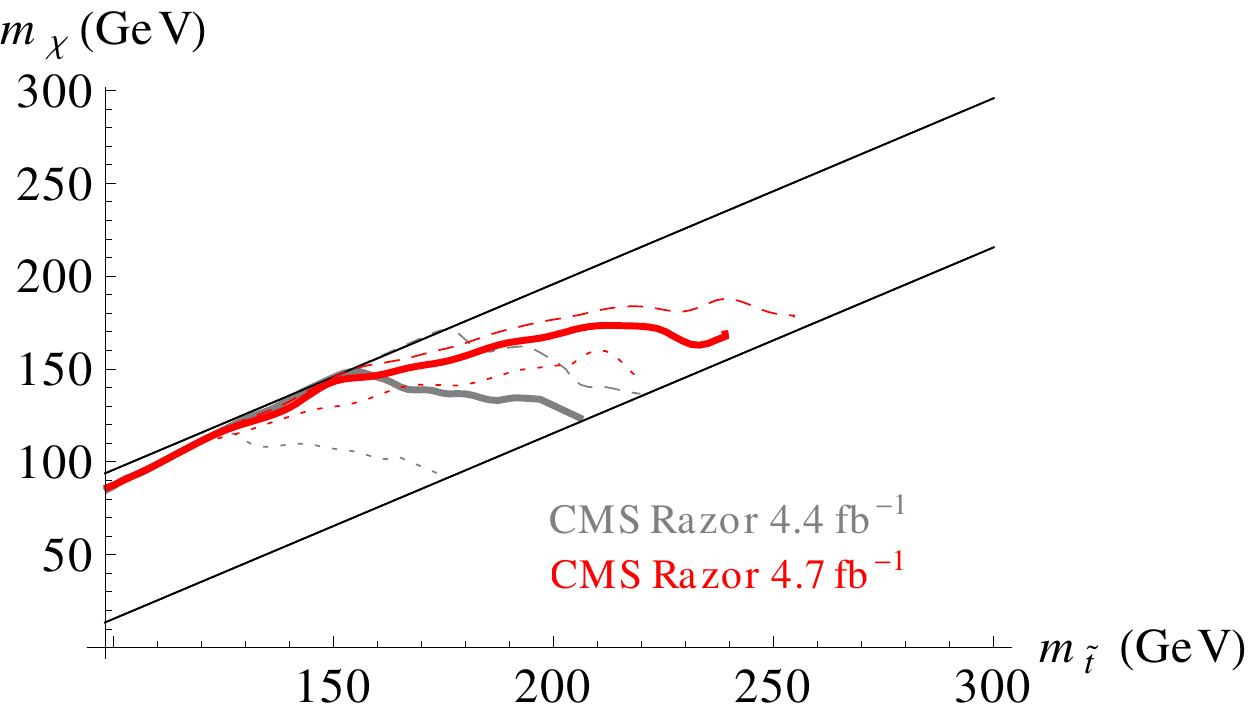}
\includegraphics[width=0.47\textwidth]{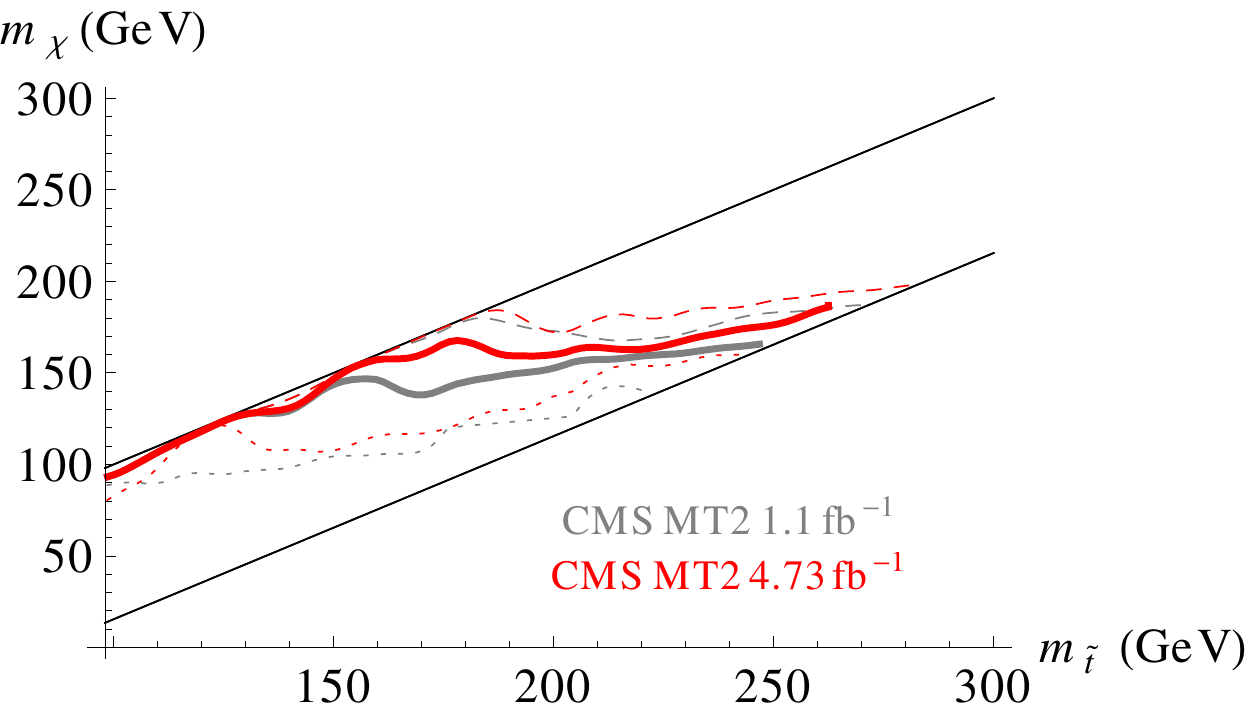}
\includegraphics[width=0.47\textwidth]{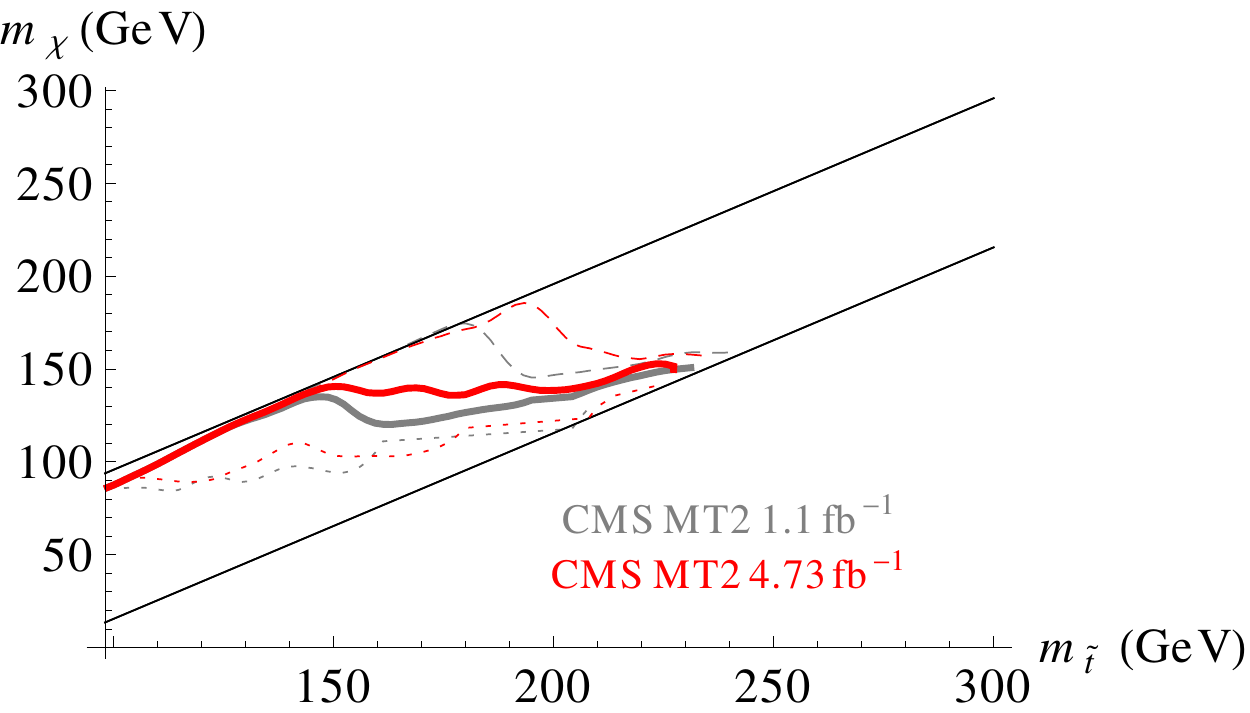}
\caption{ATLAS and CMS exclusions on FV mode~(left) and 4B mode~(right).  
The two straight, diagonal lines correspond to $\mstop=\mneut$~(upper, FV),
$\mstop=\mneut+m_b$~(upper, 4B), and $\mstop=\mneut+m_b+m_W$~(lower, both). 
The dashed/dotted lines correspond to varying the cross section by $\pm50\%$. 
The upper straight lines are the kinematic limits for these modes, 
while below the lower straight line we expect the 3B channel to take over.}
\label{fig:FV4Bdecays}   
\end{center}  
\end{figure}

\clearpage

\subsubsection{4B mode}

  Each 4B stop decay produces the final state
$\tilde{t}\to b\,f\,\bar{f}'\chi_1^0$, where the $f\bar{f}'$ fermions 
come from the decay of an off-shell $W^+$.  Events with a pair of
stops can therefore produce zero, one, or two leptons.
While the potential presence of leptons might seem promising,
we find that they tend to be somewhat soft and difficult to identify
when produced by this decay mode.\footnote{The requirement of multiple
hard jets in these analyses is also problematic.}  
Hadronic analyses turn out to be more effective, although their 
reach is slightly weaker than for the FV decay channel.
Let us also mention that 4B decays were performed in 
Pythia~6.4~\cite{Sjostrand:2006za} which uses a flat matrix element
and may be a poor approximation near kinematic boundaries.
We defer an attempt to implement the full decay matrix element to a 
future study.

  Among the primarily hadronic search channels, we find that the ATLAS
and CMS monojet and jets+$\MET$ analyses are moderately effective.
The monojet searches are sensitive mainly to the case where the hard jet 
is produced by QCD radiation, and as a result the 4B region excluded 
by the monojet searches is similar to the FV case.  The ATLAS jets+$\MET$ search
of Ref.~\cite{A2} requiring two hard jets is effective for probing
the small $(m_{\tilde{t}}-m_{\chi_1^0})$ region.  However, the higher number 
of final states in the 4B decay results in softer jets relative
to FV decays and a smaller exclusion region.  We find a similar
result for the CMS jets+$\MET$ search of Ref.~\cite{:2012mfa}.

  The CMS Razor analyses \cite{C5,C15} constrain the hadronic 4B mode 
only in the hadronic ($\emph HAD$) channel. Although some of the Razor 
analyses also search for one or two $e$ or $\mu$ in the final state, 
the leptonic 4B channel is not picked up by the search. 
The light stop decays are constrained by the low
transverse mass search region, $M_{R}<1000\gev$, and as a result, 
the excluded region is similar to the FV mode. The more recent search 
of Ref.~\cite{C15} has an additional $b$-tag requirement and more data,
and excludes a larger region.\footnote{The result of Ref.~\cite{Delgado:2012eu}, 
that appeared after the original version of this work was submitted, 
suggests an even larger exclusion is possible if the $b$-tag requirement 
is relaxed.}
Similarly, the CMS $M_{T2}$ analyses of Refs.~\cite{C7,:2012jx}
that demand a $b$-tag are also effective.  However, the net exclusions
of these analyses are slightly weaker than for the FV mode,
due to the higher multiplicity and softer momenta of the 4B stop
decay products.
Searches for $b$-jets and missing energy at CMS (with a lepton veto) 
are also effective at probing the 4B decay~\cite{C4,C14}.

\subsubsection{3B mode}

 The 3B mode is sensitive to analyses with $b$-jet final states, 
in particular the CMS $b$-jet, Razor, $M_{T2}$ analyses.
The newer CMS $b$-jet search \cite{C14} constrains the 3B mode, 
with an exclusion region comparable to the FV and 4B modes. 
In particular, the signal region with one loose $b$-jet, $1BL$, 
is most sensitive to the decay mode. 

  The CMS Razor analyses \cite{C5,C15} are sensitive to the 3B mode 
in the hadronic search channel ($\emph HAD$), with stronger limits 
from the more recent search. The decays are constrained by the 
transverse mass search region, $M_{R}<1000\,\gev$.
The CMS $M_{T2}$ analyses \cite{C7,:2012jx} provide the most stringent 
constraints,  up to $\mstop \simeq 300\,\gev$.  In particular, 
the $b$-tag signal region is most sensitive to the 
3B decay mode.

\begin{figure}[ttt]
\begin{center}  
\includegraphics[width=0.47\textwidth]{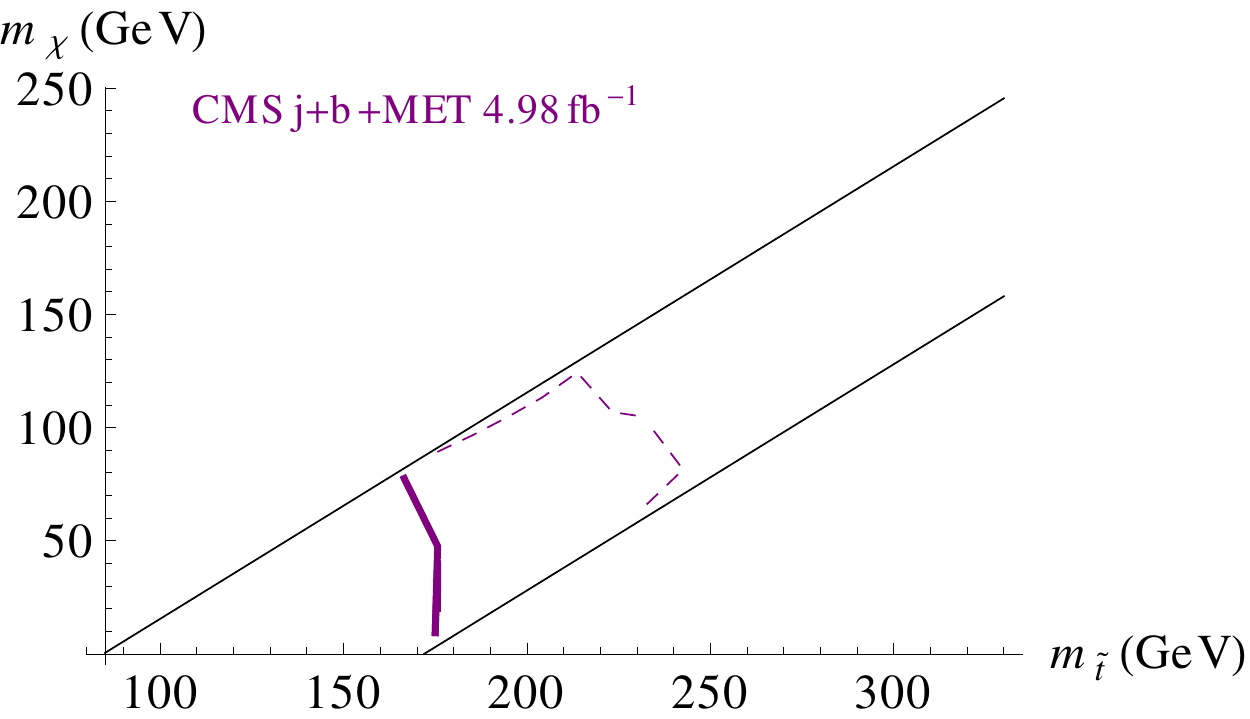}
\includegraphics[width=0.47\textwidth]{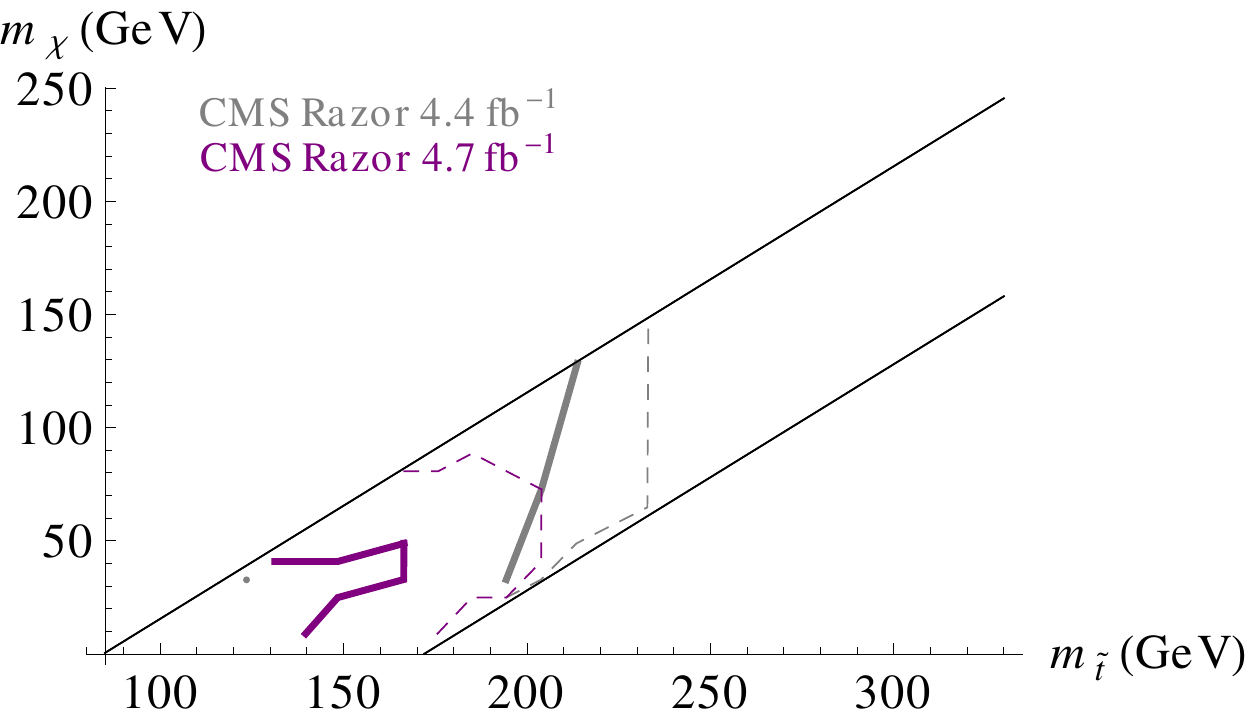}
\includegraphics[width=0.47\textwidth]{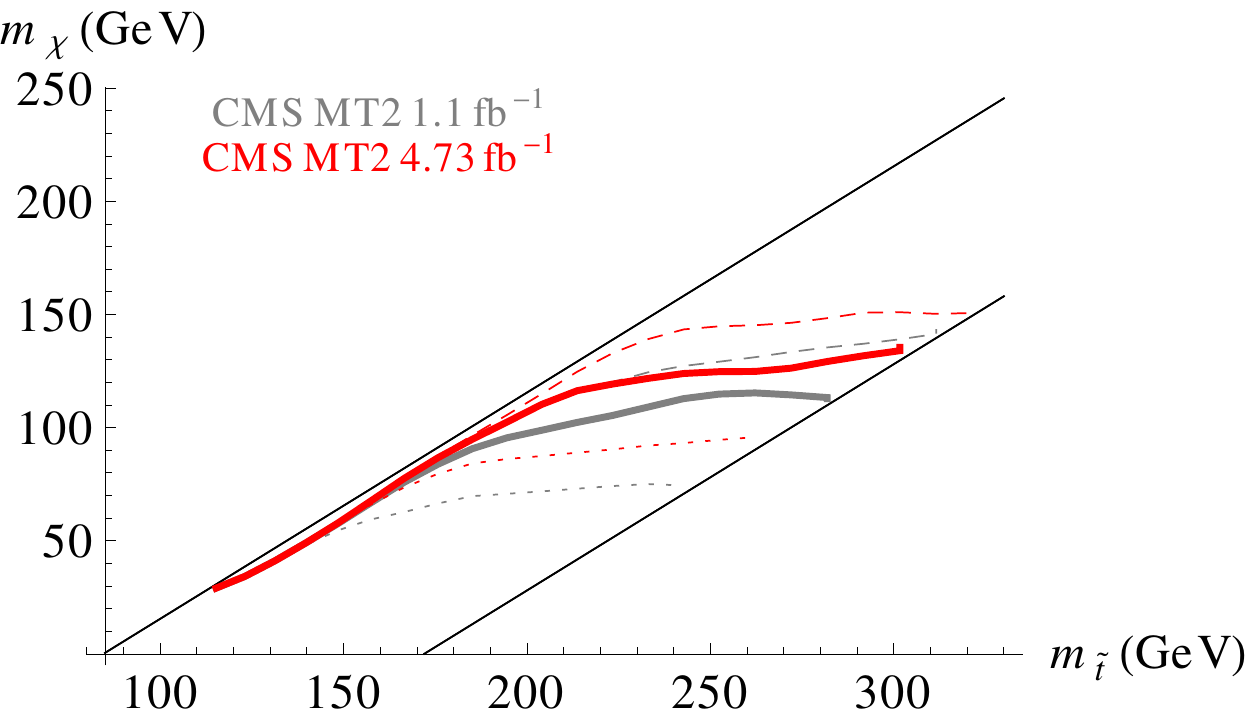}
\caption{
 CMS exclusions on 3B mode. The dashed/dotted lines correspond 
to varying the cross section by $\pm50\%$.
The upper solid diagonal line shows 
 the kinematic limit for this decay, $\mstop=\mneut+m_{b}+m_{W}$,
 while the lower solid diagonal line has $\mstop=\mneut+m_t$ where
 the 2B decay channel is expected to take over.
}
\label{fig:3Bdecays}   
\end{center}  
\end{figure}

\clearpage

\subsubsection{Summary}

  In these analyses we have only applied existing LHC searches 
using up to $5\,\text{fb}^{-1}$ of data at $\sqrt{s} = 7\,\tev$.  Extending
these analyses to the full $20\,\text{fb}^{-1}$ dataset at $8\,\tev$ will 
improve the statistical uncertainties related to these searches by 
over a factor two.  Since the signal-to-background ratios in
most of the exclusion channels considered are typically of order unity,
we expect that this to significantly improve the limits on the
light stop parameter space.  

  Even larger improvements are likely possible in these channels
with dedicated search techniques.  Many of the jet and $\met$ requirements
in the analyses applied here were designed for much heavier superpartners,
and the resulting signal rates after cuts were often very low.  
The FV search could benefit from charm tagging, which would also point 
towards a heavy-flavour origin of the jet.  In the 3B and 4B channels,
additional kinematic variables such as a $b$-lepton invariant mass
might help to isolate the signal from background and identify 
its origin~\cite{Chou:1999zb}.\footnote{We thank the referee for
suggesting this method.}

\section{Conclusion\label{sec:conc}}

  Very light stops can be consistent with current experimental bounds
when they decay in ways that are challenging to detect at the LHC.  
In this study we have attempted to investigate the properties of 
light stops in a systematic way, and to estimate the bounds placed on
them by indirect precision measurements and existing searches at the LHC.

  When a stop is very light, it can decay according to 
$\tilde{t}\to c\chi_1^0$~(FV) and $\tilde{t}\to bf\bar{f}'\chi_1^0$~(4B).
The relative branching fractions depend on the gauge content of the light
stop, the value of $\tan\beta$, and the amount of flavour mixing
within the soft parameters.  When the flavour mixing is governed by
MFV, it is possible for both the FV or 4B channels to be the dominant
decay mode.  Thus, it is important to study the signals of both possibilities
at the LHC.

  We have investigated the limits placed by existing LHC searches
at $\sqrt{s}=7\,\tev$ on a light stop decaying entirely in the FV, 4B,
or 3B channels.  Our results suggest that these searches rule out a 
light stop decaying in these channels unless it is heavier than at least
$m_{\tilde{t}} > 200\,\gev$.  Even so, we emphasize that our results
were obtained with an approximate simulation of the ATLAS and CMS detectors

  The exclusions we obtain also imply a significant bound on electroweak
baryogenesis within the MSSM.  To successfully create the observed
baryon asymmetry, a very light, mostly unmixed stop is required with 
a mass less than that of the top quark.  Our results suggest that 
such a state is ruled out by existing LHC analyses, at least if it
decays promptly in the FV, 4B, or 3B modes.  Nevertheless, extensions
of the MSSM that open up new decay channels might permit the stop
to be as light as needed.

\section*{Acknowledgements}

We thank Wolfgang Altmanshoffer, Heather Logan, Stephen Martin, 
Olivier Mattelaer, Jessie Shelton, and Chiu-Tien Yu for useful conversations.  
Johan Alwall also provided us with invaluable help with MadGraph, 
Ben Allanach aided us with SoftSUSY,
and Nazila Mahmoudi helped us with SuperISO.  DM would also like to acknowledge
the hospitality of the Perimeter Institute and the Aspen Center for Physics
while this paper was under way.  This research is supported by NSERC.

\begin{appendix}

\section{MFV Conventions and Soft Parameters\label{sec:mfv}}

  In this appendix we describe our detailed conventions for the Yukawa
couplings and the soft scalar parameters under the MFV hypothesis.
We also describe how to match them to those used in 
the SLHA2~\cite{Allanach:2008qq}
convention for specifying supersymmetric parameters.

   We define the Yukawa matrices appearing in the superpotential in
the same way as Ref.~\cite{Hiller:2008wp} (which follows 
Ref.~\cite{D'Ambrosio:2002ex}):
\beq
W \supset U^c_a(\yu^{\dag})^{a}_{~i}Q^i\cdot H_u 
- D^c_p(\yd^{\dagger})^p_{~i}Q^i\cdot H_d \ , 
\label{wmfv}
\eeq
where we have written out the explicit 
$SU(3)_Q\times SU(2)_{u_R}\times SU(3)_{d_R}$ indices.  
The relevant soft terms in the Lagrangian are
\beq
-\mathscr{L}_{soft} &\supset&
\tilde{Q}^{*}_i(m_Q^2)^i_{~j}\tilde{Q}^j
+\tilde{U}^{c}_a(m_{U^c}^2)^a_{~b}(\tilde{U}^{c*})^b
+\tilde{D}^{c}_p(m_{D^c}^2)^p_{~q}(\tilde{D}^{c*})^q\nnmb\\
&& +\tilde{U}^c_a(\Au^{\dagger})^a_{~i}Q^i\ccdot H_u
-\tilde{D}^c_p(\Ad^{\dagger})^p_{~i}Q^i\ccdot H_d \ .
\label{lsoftmfv}
\eeq
Both the superpotential of Eq.~\eqref{wmfv} and the soft terms
of of Eq.~\eqref{lsoftmfv} are invariant under the 
$G_{f} = SU(3)_Q\times SU(3)_{u_R}\times SU(3)_{d_R}$ flavour group
with the supermultiplets transforming as
\beq
Q = (3,1,1),~~~U^c = (1,\bar{3},1),~~~D^c = (1,1,\bar{3}) \ ,
\eeq
together with the Yukawa matrices transforming according to
\beq
\yu = (3,\bar{3},1),~~~~~~\yd=(3,1,\bar{3}) \ .
\eeq

  Suppose we make a field redefinition by a $G_f$ transformation
(keeping the Yukawa matrices held fixed):
\beq
Q \to V_QQ,~~~~U^c\to U_u^*U^c,~~~~~D^c\to U_d^*D^c \ .
\eeq
This modifies the effective Yukawa couplings appearing in Eq.~\eqref{wmfv}
by 
\beq
\yu^{\dagger}\to \yu^{'\dagger} = U_u^{\dagger}\yu^{\dag}V_Q,~~~~~
\yd^{\dagger}\to \yd^{'\dagger} = U_d^{\dagger}\yd^{\dag}V_Q \ .
\eeq
In this picture, the MFV hypothesis for the soft masses is
equivalent to the statement that the form of the soft terms
given in Eqs.~(\ref{mfv1}--\ref{mfv5}) remains unchanged under 
such a field redefinition, 
provided the old Yukawa couplings are replaced by the new ones
(\emph{i.e.} $\yu\to \yu'$, $\yd\to \yd'$ in Eqs.~(\ref{mfv1}--\ref{mfv5}).

  One of the two quark Yukawa matrices can be diagonalized
by making $G_f$ redefinitions.  To diagonalize both simultaneously,
we have to make a $G_f$ transformation together with separate redefinitions
for the two components of the $Q =(u_L,d_L)^t$ supermultiplet,
\beq
u_L \to V_uu_L,~~~~~d_L \to V_dd_L \ .
\eeq
This gives
\beq
Y_u^{\dagger} &\to& U_u^{\dagger}Y_u^{\dagger}V_u = \lambda_u\\
Y_d^{\dagger} &\to& U_d^{\dagger}Y_d^{\dagger}V_d = \lambda_d \ ,
\eeq
where $\lambda_u=m_u/v_u$ and $\lambda_d=m_d/v_d$ are diagonal with
real, positive entries.
In terms of the transformation matrices, the CKM matrix $V$ is found to be
\beq
V = V_u^{\dagger}V_d \ .
\eeq
This choice of field coordinates is called the super-CKM basis.
Note that these combined transformations do change the form of the effective
soft terms relative to the MFV forms of Eqs.~(\ref{mfv1}--\ref{mfv5}).

  To implement MFV soft terms, it is convenient to use $G_f$ to
choose a field basis where one of the two quark Yukawa matrices
is diagonal before applying $V_u$ and $V_d$.  
The first and most popular choice is
\beq
\yd = \lambda_d,~~~~~\yu = V^{\dagger}\lambda_u \ ,
\label{ydbasis}
\eeq
where $V$ is again the CKM quark mixing matrix.  This is a nice
choice because we only need to transform further by 
$V_u = V^{\dagger}$ (and $V_d=1$) to diagonalize the quark masses.
We use a second, similar choice in the discussion of squarks 
in Section~\ref{sec:stop},
\beq
\yd = V\lambda_d,~~~~~\yu = \lambda_u \ .
\label{yubasis}
\eeq
The additional transformation to diagonalize the quark masses
is now $V_d = V_{CKM}$ (and $V_u=1$).

 Non-diagonal scalar soft terms can be accommodated in the 
SLHA2 convention~\cite{Allanach:2008qq}.
the down-type squark mass matrix in the super-CKM basis is
\beq
\mathcal{M}_d^2 = \left(\begin{array}{cc}
\hat{m}_{\tilde{Q}}^2 + m_d^2+D_{d_L}&v_d\hat{T}_d^{\dagger}-\mu m_d\tan\beta\\
v_d\hat{T}_d - \mu^*m_d\tan\beta.&m_{\tilde{D}}^2+m_d^2+D_{d_R}
\end{array}
\right) \ ,
\eeq
while the up-type squark matrix is
\beq
\mathcal{M}_u^2 = \left(\begin{array}{cc}
V\hat{m}_{\tilde{Q}}^2V^{\dagger} + m_u^2+D_{u_L}&
v_u\hat{T}_u^{\dagger}-\mu m_u\cot\beta.\\
v_u\hat{T}_u - \mu^*m_u\cot\beta.&m_{\tilde{U}}^2+m_u^2+D_{u_R}
\end{array}
\right) \ .
\eeq
Working in a $G_f$ basis where either $\yu$ or $\yd$ is diagonal,
these entries are related to the MFV soft masses using our
conventions by
\beq
\hat{m}_{\tilde{Q}}^2 &=& V_d^{\dagger}m_{{Q}}^2V_d
=\tilde{m}^2(a_1\mathbb{I}+b_1V^{\dag}\lambda_u^2V+b_2\lambda_d^2
+ \ldots )
\\
\hat{m}_{\tilde{U}}^2 &=& m_{U^c}^2
= \tilde{m}^2(a_2\mathbb{I}+b_5\lambda_u^2+c_1\lambda_uV\lambda_d^2V\lambda_u)
\\
\hat{m}_{\tilde{D}}^2&=& m_{D^c}^2
=\tilde{m}^2(a_3\mathbb{I}+b_6\lambda_d^2 )\\
\hat{T}_u &=& a_u^{\dagger}V_u = A\lambda_u(a_4\mathbb{I}
+b_7V\lambda_d^2V^{\dagger})
\\
\hat{T}_d &=& a_d^{\dagger}V_d 
=A\lambda_d(a_5\mathbb{I}+b_8V\lambda_u^2V^{\dagger}) 
\ .
\eeq
These expressions apply for either $\yu$ or $\yd$ diagonal.

\end{appendix}

\bibliographystyle{JHEP}
\bibliography{stops-refs}




\end{document}